%% file: spectrum_NF0.tex
\documentclass[11pt,a4paper]{article}
\pdfoutput=1
\usepackage{jheppub}  
\usepackage{amsfonts,amsbsy,amsmath,amssymb}
\usepackage{color}
\usepackage{graphicx}
\usepackage[colorinlistoftodos]{todonotes}
\usepackage{booktabs}
\usepackage{caption}
\usepackage{subcaption} 
\usepackage{multirow} 
\usepackage{algorithm}
\usepackage[noend]{algpseudocode}
\usepackage[section]{placeins}
\usepackage[normalem]{ulem}

\newcommand{\comment}[1]{}

\newcommand{\Tr}{\text{Tr}}
\newcommand{\tmin}{t_{\mathrm{min}}}
\newcommand{\tmax}{t_{\mathrm{max}}}

\newcommand{\be}{\begin{equation}}
\newcommand{\ee}{\end{equation}}
\newcommand{\ba}{\begin{array}}
\newcommand{\ea}{\end{array}}
\newcommand{\baa}{\begin{array}}
\newcommand{\eaa}{\end{array}}
\newcommand{\bea}{\begin{eqnarray}}
\newcommand{\eea}{\end{eqnarray}}
\newcommand{\half}{\frac{1}{2}}

\newcommand{\hL}{\hat{L}}

\newcommand{\II}{\mathbb{I}}

\newcommand{\W}{\mathcal{W}}

\newcommand{\MP}{M_{\mathrm{PCAC}}}
\newcommand{\DWD}{D_{\mathrm{WD}}}
\newcommand{\DTM}{D_{\mathrm{TM}}}
\newcommand{\bchi}{\bar \chi}
\input{codes.tex}

\title{Meson  spectrum in the large $N$ limit.}

\author[a]{Margarita Garc\'{i}a P\'erez,}
\author[a,b]{Antonio Gonz\'alez-Arroyo,}
\author[c]{\\ Masanori Okawa}

\affiliation[a]{Instituto de F\'{i}sica Te\'orica UAM-CSIC, Nicol\'as
  Cabrera 13-15, Universidad Aut\'onoma de Madrid, Cantoblanco, E-28049 Madrid, Spain}
\affiliation[b]{Departamento de F\'{i}sica Te\'orica, 
M\'odulo 15,  Universidad Aut\'onoma de Madrid, Cantoblanco, E-28049 Madrid, Spain}
\affiliation[c]{Graduate School of Advanced Science and Engineering, Hiroshima University,
  Higashi-Hiroshima, Hiroshima 739-8526, Japan}

\emailAdd{margarita.garcia@uam.es}
\emailAdd{antonio.gonzalez-arroyo@uam.es}
\emailAdd{okawa@hiroshima-u.ac.jp}

\abstract{ 
    We present the result of our computation of the lowest lying meson masses for SU(N) gauge  theory
    in the large $N$ limit (with $N_f/N\longrightarrow 0$). The final values are given in units of the
    square root of the string tension, and with errors which account for both statistical and systematic
    errors. By using 4 different values of the lattice spacing we have seen that our results scale properly.
    We have studied various values of $N$ (169, 289 and 361) to monitor the N-dependence of the most sensitive
    quantities. Our methodology is based upon a first principles approach (lattice gauge theory) combined with
    large $N$ volume independence. We employed both  Wilson fermions and twisted mass fermions with maximal
    twist.  In addition to masses in the pseudoscalar, vector, scalar and axial vector channels, we also give
    results on the pseudoscalar  decay constant and various remormalization factors. 
}

\preprint{%
{
IFT-UAM/CSIC-20-110, HUPD-2003, FTUAM-20-13
}}

\begin{document}


\maketitle

\section{Introduction}
\label{s:introduction}
\input{introduction.tex}

\section{Meson masses from reduced models}
\label{s:main_formulas}
\input{main_formulas.tex}

\section{Methodological aspects}
\label{s:methodology}
\input{data.tex}

\subsection{Extraction of masses}
\input{mass_computation.tex}

\subsection{Analysis of errors: statistical and systematic}
\label{s:errors}
\input{errors.tex}
\section{Results and analysis}
\label{s:results}
\subsection{Pseudoscalar channel}
\input{pion_channel.tex}

\subsection{The vector channel}
\input{vector_channel.tex}
\subsection{Other quantum numbers}
\input{heavy_states.tex}

\section{Summary of Results and Discussion}
\label{s:discussion}
\input{discussion.tex}
\section*{Acknowledgments}
We acknowledge interesting conversations about various aspects with G. Bali, M. Bocchicchio, B. Lucini, C. Pena and F. Romero-L\'opez.  
M.G.P. and A.G-A acknowledge financial support from the MINECO/FEDER grant
FPA2015-68541-P  and PGC2018- 094857-B-I00 and the MINECO Centro de Excelencia Severo Ochoa Program SEV-2016-0597.
M. O. is supported by JSPS KAKENHI Grant Number 17K05417.  This publication is supported by the European project H2020-MSCAITN-2018-813942 (EuroPLEx) and the EU Horizon 2020 research and innovation programme, STRONG-2020 project, under grant agreement No 824093.
This research used computational resources of the SX-ACE system provided by
Osaka University through the HPCI System Research Project (Project ID: hp170003 and hp180002).
We acknowledge the use of the Hydra cluster at IFT.

\appendix 
\section{Raw data}
\label{a:tables}
\input{tables.tex}
\FloatBarrier

\input{spectrum_NF0.bbl}

\end{document}

%% file: codes.tex
\newcommand{\MESONS}{Gonzalez-Arroyo:2015bya}
\newcommand{ \TESTING}{Gonzalez-Arroyo:2014dua}
\newcommand{ \STRTENSNEW}{GonzalezArroyo:2012fx}
\newcommand{ \TEKFOUR}{GonzalezArroyo:2010ss}

\newcommand{ \STRTENSOLD}{GonzalezArroyo:1983pw}
\newcommand{ \TEKTHREE}{Aldazabal:1983ec}
\newcommand{ \TEKTWO}{GonzalezArroyo:1982hz}
\newcommand{ \TEKONE}{GonzalezArroyo:1982ub}
\newcommand{\TWISTTHREE}{tHooft:1981sps}
\newcommand{\TWISTTWO }{tHooft:1980kjq}
\newcommand{\TWISTONE}{tHooft:1979rtg}

\newcommand{ \EK}{Eguchi:1982nm}
\newcommand{\VOLINDTHREE}{Kovtun:2007py}
\newcommand{\VOLINDFOUR}{Unsal:2008ch}
\newcommand{\NARNEUREDCONT}{Narayanan:2003fc}

\newcommand{ \BALIMESONS}{Bali:2013kia}

\newcommand{ \DEGRANDMESONS }{DeGrand:2016pur}
\newcommand{\LUCTEPTHREE}{Lucini:2002ku}
\newcommand{ \LUCTEPTWO}{Lucini:2001ej}
\newcommand{\QEK}{Bhanot:1982sh}
\newcommand{\GEVPONE}{Michael:1985ne}
\newcommand{\GEVPTWO}{Luscher:1990ck}

%% file: introduction.tex
Large $N$ gauge theories~\cite{thooft:1973alw} sit  at the crux of different approaches to
quantum field theory. This is one of the motivations for studying
these theories. For the most basic case of a pure Yang-Mills theory we
face the difficulties of a strong non-perturbative dynamics without
the help of supersymmetry. Nonetheless, the theory exhibits many
simplifications with respect to theories at finite $N$. This comes
without paying the price of loosing some of the main non-perturbative
phenomena, such as confinement or existence of a mass gap, which
remain a challenge for a rigorous proof. From a physicist standpoint
it is important to extract the values of the main physical parameters 
of the theory. These values  will remain a  testing ground for new
ideas and methodologies aiming at an analytic or numerical approach to
the theory. In particular, as mentioned earlier, the large $N$ theory seems
more easily addressable from studies originating from string theory,
as the AdS/CFT approach~\cite{Maldacena:1997re,Gubser:1998bc,Witten:1998qj}.   The main  first principles
approach to non-perturbative dynamics is given by Wilson's lattice
gauge theories~\cite{Wilson:1974sk}. Thus, it is extremely desirable to attack the problem
using this methodology. The importance of this goal has been realized
from long time ago by a small fraction of the community, pioneered by
the work of Mike Teper and collaborators~\cite{\LUCTEPTWO, \LUCTEPTHREE } (See Ref.~\cite{Lucini:2012gg} for a review of results). The intrinsic
difficulty of this goal comes from the fact that the large $N$ results
follow from extrapolation of those obtained at several small values of
$N$. Fortunately for this program, it seems that many of the physical
quantities have a relatively mild dependence on $N$. On the negative side it turns out that the large $N$ gauge theories possess simplifications that are not present in theories at finite  $N$, making the study of the latter harder.  To name one,  which is of particular relevance for this work, we have the role paid by quark degrees of freedom in the fundamental representation. In the
large $N$ limit these quarks are non-dynamical, so that the so-called
quenched approximation becomes exact. On the other hand, at finite $N$
fundamental quarks are dynamical. Generating the configurations
becomes much more costly,  and if the  quenched approximation is used 
one has to be careful in treating the chiral behaviour of the theory. 
There is certainly an interplay between the small quark mass and large
$N$ limits that one has to be careful about~\cite{Kaiser:2000gs, Kaiser:1998ds}.

The present work follows a completely different approach. The idea is
to exploit one of the simplifications that the large $N$ limit
produces. In particular, we will employ the property of volume
independence, originating from an observation  by Eguchi and
Kawai~\cite{\EK}. In essence, the statement says that finite
volume corrections go to zero in the large $N$ limit. Although the
original formulation was proven incorrect in the continuum
limit~\cite{Bhanot:1982sh}, several ways to enforce the property have been
found~\cite{\QEK,\NARNEUREDCONT,\VOLINDTHREE, \VOLINDFOUR}. Here we will make use of an idea originating soon
after the work of Eguchi and Kawai by two of the present
authors~\cite{\TEKONE, \TEKTWO}. The main   point is to use appropriately
chosen twisted boundary conditions~\cite{\TWISTONE , \TWISTTWO ,
\TWISTTHREE }. In this way one
can preserve the necessary symmetries to guarantee that the original
proof of Eguchi and Kawai holds. As a matter of fact one can take
volume independence to the extreme and reduce the lattice volume to
single point. This gives a matrix model of $d=4$ matrices known as the
Twisted Eguchi-Kawai model (TEK). The advantage of this formulation is
that because of the reduction in the number of space-time degrees of
freedom, one can go to much larger values of $N$. Thus large $N$
extrapolations become unnecessary and some corrections like those
associated to quark loops become negligibly small. There are, however,
finite $N$ corrections which turn out to be of a different nature than
those of the ordinary theory. These corrections are of two types. One
takes the form of ordinary finite volume lattice corrections on a
lattice of size $(\sqrt{N})^4$. Hence, a fixed value of $N$ determines a scaling window in which the effective size of the box in physical units remains large
enough. Going to smaller lattice spacings one would enter the
femtoworld like in standard lattice gauge theories. To enlarge the window one has to increase the value of $N$. The other type of
correction is related to an incomplete cancellation of non-planar
contributions. The nature of these corrections are related to effects
in non-commutative field
theory~\cite{GonzalezArroyo:1983ac, Connes:1987ue, Douglas:2001ba}. Fortunately,
the formulation of TEK has a free integer
flux parameter that can be tuned to minimize these corrections and
used to quantify their effect~\cite{\TEKFOUR}. 

In summary, our formulation of the large $N$ lattice theory (the TEK
model) has the capacity of producing estimates of the physical
observables of the theory with all statistical and systematic errors
under control: measurable and improvable. In the last few years we
have been running a number of tests~\cite{\TESTING} to ensure the capacity of our method to produce physical results within the standard computational
resources available at present. The first continuum observable that was studied 
was the string tension. Indeed, the early estimates~\cite{\STRTENSOLD,Fabricius:1984un} 
using our method preceded in many years any other large $N$ lattice estimate. 
More recently~\cite{\STRTENSNEW}, a measurement to a few percent precision
has been obtained, which is at least of the level of precision as
other recent determinations~\cite{Lohmayer:2012ue,Athenodorou:2010cs}

The present work aims at producing a calculation of the low-lying meson spectrum in the large $N$ limit with all statistical and systematic errors under
control. The work is the result of several  years of study. The
methodology was developed in Ref.~\cite{\MESONS}. Several
tests and technical improvements, as well as partial results  have been 
presented in  other works~\cite{Gonzalez-Arroyo:2015chm,Perez:2016zjk,
Perez:2020fqn}. The meson spectrum has also been investigated by other
authors using different techniques. Indeed, this  was pioneered by
extrapolations from quenched
studies~\cite{DelDebbio:2007wk,Bali:2008an}. There are also studies
which employed the idea of volume independence to give
determinations~\cite{Hietanen:2009tu}. More recent determinations 
based on the extrapolation by other authors~\cite{\BALIMESONS,\DEGRANDMESONS,Hernandez:2019qed}
give  results with  similar precision to ours. The exact compatibility demands that all estimates have well quantified errors. This is our purpose here. It must be said
that the existence of different methodologies is very welcome since
there are obviously pros and cons in each of them. We have already
mentioned some advantages of our method in relation with the quenched
approximation and the chiral limit. However, on the negative side
our method does not allow the computation of $1/N$ corrections, which
are also important phenomenologically. Combining our results with
finite $N$ estimates could be very interesting. 

The structure of the paper is as follows. In the next section we review the philosophy 
and main formulas involved in our methodology, referring to the original references for explicit derivations.
Section~\ref{s:methodology} is of the most technical nature. Readers mostly interested in the results might skip it. However, for us this section is crucial since it lists and analyzes all the steps involved in giving final numbers and the possible sources of errors they might introduce. A lot of effort has been put into it so that our final mass values have trustworthy statistical and systematic errors. Section~\ref{s:results} contains the presentation of our results starting with those associated with the pseudoscalar channel. In this  channel we use both Wilson and maximally twisted mass fermions. This turns out to be crucial to obtain a determination of the pseudoscalar decay constant. The vector, scalar, and axial vector meson masses are computed as well. In Section~\ref{s:discussion} we use the previous information to present our final table of values of the meson  masses in the continuum limit. Values are presented with statistical errors and systematic errors mainly arising from the extrapolation to vanishing lattice spacing. Readers interested in the results should address directly to this section. We also compare our results with other  determinations and predictions of the meson masses and the pion decay constant.   Possible future improvements are discussed.  At the end of the paper we give a long list of tables containing the explicit lattice results of our simulations. This might be useful for other researchers who might want to use our bare results for comparison or display. We have ourselves profited from other authors doing the same.

%% file: main_formulas.tex
As explained in the introduction, the goal of this paper is to compute the masses of mesons with small
number of flavours in the large $N$ limit of $d=4$ Yang-Mills theory. Our
method is based on reduced models. In particular we will be using the
twisted Eguchi-Kawai model, which is a model involving $d$ SU(N)
matrices without any space-time label. Despite its conceptual
simplicity this matrix model has observables whose expectation value
in the large $N$ limit coincides with that of ordinary lattice gauge
theory at infinite volume and infinite $N$. In particular this refers
to Wilson loops. This  statement can be proven  non-perturbatively
under certain assumptions by Schwinger-Dyson equations\cite{\EK,\TEKONE}, perturbatively
to all  orders~\cite{\TEKTWO,\TEKTHREE} and also tested  up to 5 decimal
places  by direct evaluation of Wilson loops in the ordinary and
reduced model~\cite{\TESTING}. 

Let us briefly revise the probability distribution and action density
of the TEK model. The former is given in terms of the latter by means
of the partition function 
\be
Z_{TEK}= \prod_{\mu=0}^{d-1} \left(\int dU_\mu\right)\  e^{-S_{TEK}}
\label{partfunc}
\ee
where the integration on the SU(N) $U_\mu$ matrices uses the invariant
Haar measure on the group. The action $S_{TEK}$ follows by contracting to a
point the action of ordinary lattice gauge theories in a finite box
with twisted boundary conditions. Many different types of lattice
actions can be used, but most of the numerical results have been
obtained using the simple Wilson action
\be
S_{TEK}= -bN \sum_{\mu\ne \nu} z_{\nu \mu} \Tr\left(U_\mu U_\nu
U_\mu^\dagger U_\nu^\dagger\right)
\label{TEKaction}
\ee
where $1/b$ is the lattice equivalent of `t Hooft coupling $\lambda$, and the
factors $z_{\nu \mu}=z^*_{\mu \nu}=\exp\{2 \pi i n_{\nu \mu}/N\}$ are Nth 
roots of unity. The integer-valued antisymmetric twist tensor $n_{\nu
\mu}$ is irrelevant in the large $N$ limit, provided it is chosen in
the appropriate range. A bad choice can yield very large finite $N$
corrections and even a possible breaking of the symmetry conditions
for reduction to hold~\cite{ishikawa:2003, Bietenholz:2006cz,
Teper:2006sp}. In practice our choice has been  the  so-called 
symmetric twist which demands $N$ to be the square of an integer
$N=\hat{L}^2$. Then one takes $n_{\nu \mu}=k\hat{L}$ for $\mu>\nu$,
where $k$ is an integer coprime with $\hat{L}$.  The choice is irrelevant provided one satisfies the criteria given in Ref.~\cite{\TEKTHREE}. 

Reduction implies that the expectation value of a Wilson loop
$W(\mathcal{C})$ for an  SU(N) gauge theory at infinite volume and
infinite $N$ can be obtained as follows
\be
\label{loops}
W(\mathcal{C})= \lim_{N\longrightarrow \infty}
\frac{z(\mathcal{C})}{N}\langle \Tr( U(\mathcal{C})) \rangle_{TEK}
\ee
where the right-hand side is an expectation value in the TEK model
of the trace of the product of link variables following the sequence
of directions specified by $\mathcal{C}$ (no space-time labels for
TEK links). The factor $z(\mathcal{C})$ is a product of the $z_{\nu
\mu}$ factor for a collection of plaquettes with boundary on
$\mathcal{C}$. 

The previous formula \eqref{loops} can be proven non-perturbatively
from Schwinger-Dyson equations assuming center symmetry~\cite{\EK,\TEKONE},  shown to 
be valid in perturbation theory to all orders~\cite{\TEKTWO}, and checked
directly to very high precision numerically ~\cite{\TESTING}.
Furthermore, these results give  information about how big has $N$ to
be to acquire a given precision. Indeed, part of the finite $N$
corrections assume the form of finite size corrections for a lattice
of size $(\sqrt{N})^4= \hat{L}^4$. Hence, when computing extended observables it
is necessary that they all fit inside such an effective box.  Typical
loop sizes should then be smaller than $\hat{L}$. Lattice
masses should also satisfy $M\hat{L} \gg 1$. In practice, due to
scaling, this affects the range of values of $b$ to be covered. All
these limitations are pretty standard in all lattice gauge theory
studies. Here we simply have to look at $\hat{L}=\sqrt{N}$ as the
effective box size. The whole procedure has been tested quite
successfully in computing the string tension~\cite{\STRTENSNEW} in the
large $N$ limit  and other observables~\cite{Perez:2014isa}. In
those studies we went as far as $N=1369$, corresponding to a box size
of $37^4$, which is quite satisfactory for lattice standards.
Unfortunately, in the present study we are unable to reach those
values because of the computational effort involved in computing the
quark propagator. Our work is mostly based on results obtained for $N=289$
corresponding to a box size of $17^4$. This  limits the range of our
$b$ values, but still seems sufficient to obtain good estimates of the
corresponding masses. We have also obtained some results at
$\hat{L}=13$ and $19$, allowing us to test the effect of finite $N$.

The methodology for computing meson masses  has been explained in
previous publications of two of the present authors~\cite{\MESONS}. We
will assume that the number of flavours of quarks in the fundamental
stays finite in the large $N$  limit. Hence, quarks are non-dynamical
and the configurations are generated with the pure gauge TEK model. 
Quarks do propagate in the background of these gauge fields. One can
allow the quark fields to propagate in a lattice of any size (with
restrictions) including infinite size. In practice, since the gauge 
fields live in an effective box
of size $\hL^4$, we can restrict the quarks to live in a box of the
same size, although we will double it in the time direction for ease
in computing correlators. 

The meson masses are obtained by measuring the exponential decay in
time of correlators among quark bilinear  operators projected to vanishing
spatial momentum. For example, one can consider objects of the form
\be
\mathbf{O}(t)=\bar\Psi(t,\vec{x}) O(\vec{x},\vec{y}) \Psi(t,\vec{y}) 
\ee
where $O$ is also a matrix in spinor and colour indices. 
Correlators of these bilinears, after integration over the fermion
fields, become expectation values in the probability distribution
provided by the TEK action:
\be
\label{correlators}
C(t)=\langle \mathbf{O}(0) \mathbf{O'}(t) \rangle  =
 \langle -\Tr(
 O(\vec{x},\vec{y}) D^{-1}(0,\vec{y};t, \vec{z})  O'(\vec{z},\vec{u})
 D^{-1}(t,\vec{u};0, \vec{x})) \rangle
\ee
where $D^{-1}(x,y)$ stands for the lattice quark propagator between
two space-time points. The lattice propagator is the inverse of the
Dirac operator. On the lattice there are several options. Most of our
results are obtained for the Wilson-Dirac operator, although we will
also use the twisted mass~\cite{Frezzotti:2000nk, Frezzotti:2003ni, Shindler:2007vp} operator. Other 
possibilities are feasible but have not been used in this work. The
set of meson operators used for our work includes both local and
non-local ones and will be explained later. All of the operators are
gauge-invariant and projected to vanishing spatial momentum. It is important to use
various operators with the same quantum numbers, since the
corresponding masses should coincide. This helps in obtaining more
precise determination of the masses. Another comment is that in
Eq.~\eqref{correlators} we have omitted disconnected pieces, which are
subleading at large $N$. This is an important simplification, which
erases the difference between flavour singlet and non-singlet channels.  

The description done in the previous paragraph looks completely
standard. The main difference in our case is that quarks propagate in
the background  of volume independent gauge fields (up to a twist).  This
simplifies considerably the structure of the propagators and the meson
correlators. An analogy can be drawn with the situation of electrons
propagating in a crystal solid. The background field in that case is
the potential created by the ions of the solid, which is periodic with
a period of the order of the lattice spacing. The motion of the
electron in an infinite or much bigger solid translates into a
dependence of the propagator on the Bloch momentum. Our case is not
exactly identical since the gauge field is only periodic up to a twist. 
The gauge field only repeats itself exactly after $\sqrt{N}$ steps in
any direction. However, we only consider situations in which the
correlation lengths are much smaller than this scale. For that
purpose, obviously $N$ should be large enough. 

Following the derivation given in Ref.~\cite{\MESONS} we arrive at the
following formula for the meson correlator in momentum space
\be
\label{corr_mom}
\hat{C}(p_0)=\sum_{q} \langle \Tr(O D^{-1}(q_0, \vec{q}) O'
D^{-1}(q_0+p_0, \vec{q})) \rangle
\ee
where the trace runs over colour and spinorial indices. In a first
approximation the operators $O$ and $O'$ are just  different matrices 
of the Dirac Clifford algebra defining the quantum numbers of the
meson to be studied. Finally, $D^{-1}(q_0, \vec{q})$ is the
inverse of the Dirac operator  in our particular setting. We will
first of all  focus on Wilson fermions, on which most of our results are based.  
As explained in Ref.~\cite{\MESONS}, after some algebra one  can
write the Dirac operator as 
\be
\DWD(p)=\II -\kappa \sum_{\mu=0}^{d-1} \left( (\II +\gamma_\mu)\otimes
\W_\mu(p)+ (\II -\gamma_\mu)\otimes  \W_\mu^\dagger(p)\right)
\ee
where $\W_\mu(p)$ is the following $N^2\times N^2$ matrix
\be
\W_\mu(p)= e^{i p_\mu a} U_\mu \otimes \Gamma_\mu^*
\ee
where $U_\mu$ are the SU(N) matrices generated by the standard Monte Carlo
method for the TEK model and $\Gamma_\mu$ are fixed SU(N) matrices
satisfying 
\be
\Gamma_\mu \Gamma_\nu = e^{2 \pi i n_{\mu \nu}/N} \Gamma_\nu \Gamma_\mu
\ee
where  $n_{\mu \nu}$ is the same twist tensor used in the TEK action. 
With our choice of twist tensor the solution of the previous equation
is unique up to similarity transformations and  multiplication by
elements of the center. The first ambiguity is irrelevant since our
observables are traces. The phase ambiguity can be reabsorbed into
$U_\mu$. In practice, we can pick any particular solution and keep it
fixed.   

A final comment affects the range of values of momenta. This is determined by  the size of the lattice in which the quark fields  live. If
quarks propagate in  an infinite lattice, then $p_\mu$ is an angle.  However, since the gluon fields  
are equivalent  to those  living in a box of size $\sqrt{N}^4$, it is reasonable
to confine the fermions to live in a similar box. In that case,   momenta range over  integer multiples of $2 \pi/(a \sqrt{N})$.  This choice has a bonus, since  then 
 the momentum factors can be reabsorbed in
$U_\mu$, and one can omit the sum over $q$ in Eq.~(\ref{corr_mom}). For the total temporal momentum 
we chose $p_0=\pi n_0/(a\sqrt{N})$ with $n_0$ the integer ranging from 0 to $2\sqrt{N} -1$, allowing propagators to extend longer in the time direction.  The correlator in
configuration space is then given by 
\be
\label{corr_time}
C(t)= \frac{1}{2 \sqrt{N}} \sum_{n_0=0}^{2 \sqrt{N}-1}  e^{- i \pi t
n_0/ (a \sqrt{N})} \hat{C}(p_0)
\ee
It is this observable that has been used to extract masses of Wilson fermions.

%% file: data.tex
\subsection{Data sample and scale setting}
The analysis to be presented is based on several years of work
accumulating configurations generated with the TEK probability density
based on Wilson action as shown in Eqs.~\eqref{partfunc}-\eqref{TEKaction}. The main parameters
defining each simulation are the value of $\hat{L}=\sqrt{N}$, the
inverse `t Hooft coupling $b$, and the value of the twist flux integer
$k$. The total number of configurations accumulated for each set of
parameters are given on Table~\ref{tableconf}. For each value a number of
thermalization steps was performed initially. We do not appreciate any
Monte Carlo time dependence of our results. The configurations used
for the analysis were generated using the overrelaxation method
explained in Ref.~\cite{Perez:2015ssa}, which gives shorter autocorrelation
times that the more traditional modified heath-bath  $\acute{\rm a}$ la
Fabricius-Hahn~\cite{Fabricius:1984wp}. The number of sweeps performed from one configuration to the next $N_s$, also shown in Table~\ref{tableconf}, is chosen large to ensure that the configurations are largely independent.

\begin{table}
  \begin{center}
\begin{tabular}{|c|c|c|c|c|c|c|c|c|}\hline 
$N$ &$k$&   $b$ & $N_{\mathrm{confs}}$  & $N_s$ & $\sqrt{\sigma} a(b)$ &  $a(b)/\bar{r}$ & $a(b)/\sqrt{8 t_0}$ & $l(b,N)\sqrt{\sigma}$ \\ \hline 
169 & 5 & 0.355 & 1600 & 1000 & 0.2410(31) & 0.2389(20)  & 0.2271(1) & 3.133(40) \\ 
169 & 5 & 0.360 & 1600 & 1000 & 0.2058(25) & 0.2015(12)  & 0.1933(1) & 2.675(33) \\ \hline 

289 & 5 & 0.355 & 800  & 1000 & 0.2410(31) & 0.2389(20)  & 0.2271(1) & 4.097(53) \\ 
289 & 5 & 0.360 & 800  & 1000 & 0.2058(25) & 0.2015(12)  & 0.1933(1) & 3.499(43) \\ 
289 & 5 & 0.365 & 800  & 1000 & 0.1784(17) & 0.1722(11)  & 0.1661(1) & 3.032(29) \\ 
289 & 5 & 0.370 & 800  & 1000 & 0.1573(18) & 0.1501(10)  & 0.1434(1) & 2.674(30) \\ \hline

361 & 7 & 0.355 & 800  & 3000 & 0.2410(31) & 0.2389(20)  & 0.2271(1) & 4.579(59) \\ 
361 & 7 & 0.360 & 800  & 3000 & 0.2058(25) & 0.2015(12)  & 0.1933(1) & 3.910(48) \\ 
361 & 7 & 0.365 & 800  & 3000 & 0.1784(17) & 0.1722(11)  & 0.1661(1) & 3.389(32) \\ 
361 & 7 & 0.370 & 800  & 3000 & 0.1573(18) & 0.1501(10)  & 0.1434(1) & 2.989(34) \\ \hline
\end{tabular}
\end{center} 
\caption{Our data sample: For each value of $N$, $k$ and $b$, we list the  number of configurations $N_{\mathrm{confs}}$ , number of sweeps $N_s$ between configurations,
the lattice spacing in various units ($\sqrt{\sigma} a(b)$, $a(b)/\bar{r}$,  $a(b)/\sqrt{8 t_0}$) and the   effective box size $l(b,N)$ in $1/\sqrt{\sigma}$ units.}
\label{tableconf}
\end{table}

The choice of parameters explained in the previous paragraph  was
dictated by the standard lattice requirement of having small lattice
spacings $a(b)$ in physical units. These values can be extracted from
our previous analysis of the string tension~\cite{\STRTENSNEW},
which used much larger values of $N$ and many more values of $b$. 
Having 4 different values of $b$ in our case will allow us to test the
scaling behaviour of our data. Scale-setting is an important step in providing physical values in the continuum limit. The idea is to express all dimensionful magnitudes in terms of an appropriate physical unit. Scaling dictates that choosing one unit or other is irrelevant as we approach the critical point (in our case $b=\infty$). Apart from the string tension one can choose other physical  units. The constancy of the ratio of units as we approach the continuum limit is by itself a check of scaling. Many  proposals appear in the literature. A good unit must be easily computable and less affected by corrections such as lattice artifacts or finite volume corrections. One possible choice is given by the scale $\bar{r}$ introduced in Ref.~\cite{\STRTENSNEW}. Essentially, it is similar in spirit to Sommer scale~\cite{Sommer:1993ce} but defined in terms of square Creutz ratios. Our determination, mainly based in our extensive analysis done for our aforementioned string tension paper, is included in the table.  
A very popular scale lately is the quantity $\sqrt{8 t_0}$ defined in terms of the Wilson flow~\cite{Luscher:2010iy}. We have analyzed this quantity for the TEK model and various values of $b$ and $N$. Using this information we were able to extrapolate the value of  $\sqrt{8 t_0}$ to $N=\infty$. The results are given in Table~\ref{tableconf}. Errors come from a simultaneous fit to various $N$ and are presumably underestimated. Remarkably a simultaneous fit to the ratio $\sqrt{8 t_0}/{\bar{r}}$ and $b\ge 0.36$ gives $1.042(3)$ with $\chi^2$ smaller than 1. This is a good check of scaling given that both scales involved are based on quite different observables and procedures. Using our data for $b \ge 0.36$ we can give determinations $\sqrt{8 t_0 \sigma}=1.078(9)$ and  $\bar{r}\sqrt{\sigma}=1.035(7)$. The last number can be compared with the estimate $1.051(5)$ given in Ref.~\cite{\STRTENSNEW}, and involving data up to $b=0.385$. In conclusion, the different units used are consistent with scaling and translate into possible errors of  $\lesssim 2\%$ in the continuum limit. 

The standard lattice condition of having sufficiently large
physical volumes now translates into large enough values of $N$. We
recall that  $\sqrt{N}$ plays the role of effective size
in lattice units. Using the lattice spacing values described in the
previous paragraph one can compute the lattice effective linear size
of our data $l(b,N) \equiv \sqrt{N}a(b)$. This is given in table~\ref{tableconf} in string tension units. 

\begin{table}
  \begin{center}
\begin{tabular}{|c|c|c|c|c|c|}\hline 
meson & $\pi$ & $\rho$ & $a_0$ & $a_1$& $b_1$  \\ \hline 
$\gamma_A$ & $\gamma_5$ & $\gamma_i$ & 1 & $\gamma_5\gamma_i$ &$\frac{1}{2}\epsilon_{ijk}\gamma_j\gamma_k$ \\ \hline 
$J^{PC}$ &  $0^{-+}$ &   $1^{--}$ & $0^{++}$ & $1^{++}$ & $1^{+-}$ \\ \hline
\end{tabular}
\end{center} 
\caption{Quantum numbers of the meson channels analyzed in this work and spin-parity assignment. }
\label{meson_content}
\end{table}
\subsection{Observables}
\label{s:observables}
As explained earlier, our main observables are the meson correlation function in channel $\gamma_A$ and $\gamma_B$ at time distance $t$ 
\begin{equation}
C(t; \gamma_A, \gamma_B)=\frac{1}{2 \sqrt{N}} \sum_{n_0=0}^{2\sqrt{N}-1}{\rm e}^{-i \pi  t n_0 /(a\sqrt{N})} {\rm Tr} \left[\gamma_A D^{-1}(0)\gamma_B D^{-1}(p_0)\right]
\label{meson_prop}
\end{equation}
For the case of Wilson fermions, the Wilson-Dirac matrix $\DWD(p_0)$ depends on the
value of the hopping parameter $\kappa$ which is a function of the
bare quark mass. 
$\DWD(p_0)$ acts on color ($U_\mu$), spatial ($\Gamma_\mu$) and Dirac ($\gamma_\mu$) spaces and its explicit form is
\begin{equation}
 \DWD(p_0)=1-\kappa\sum_{\mu=0}^{d-1}\left[(1-\gamma_\mu){\tilde U_\mu} \Gamma_\mu^{*} +
(1+\gamma_\mu){\tilde U_\mu^\dagger} \Gamma_\mu^{t} \right]
\label{quark_action}
\end{equation}
with
\begin{equation}
{\tilde U_{\mu=0}}={\rm e}^{ip_0a} U_{\mu=0}={\rm e}^{i\pi n_0/\sqrt{N}} U_{\mu=0},\ \ \ \ \ {\tilde U_{\mu=1,2,3}}=U_{\mu=1,2,3} .
\label{tildeU}
\end{equation}
$\gamma_A$ and $\gamma_B$ assign meson quantum numbers in the continuum
limit. The choice of elements and their continuum spin parity
correspondence is given in Table~\ref{meson_content}.

Eq.~(\ref{meson_prop}) corresponds to correlators of  ultralocal operators.  
In order to have a range of
operators having the same quantum numbers we make use of the smearing method.
Smearing can be implemented by replacing $\gamma_A$ in Eq.~(\ref{meson_prop}) by the operator~\cite{\BALIMESONS}:
\begin{equation}
\gamma_A \rightarrow \gamma_A^\ell\equiv D_s^\ell \gamma_A,\ \ \ D_s \equiv \frac{1}{1+6c}\left[1+c \sum_{i=1}^{3}\left( {\bar U_i} \Gamma_i^{*} +
{\bar U_i^\dagger} \Gamma_i^{t} \right) \right].
\label{smearing}
\end{equation}
Here, $\ell$ is the smearing level and ${\bar U_i}$ is the APE-smeared spatial link variable obtained 
after iterating $n_{ape}$ times the following transformation
\begin{equation}
U'_i = {\rm Proj_{SU(N)}} \left[ (1-f)U_i + \frac{f}{4} \sum_{j \ne i} ( U_j U_i U_j^{\dagger} + U_j^{\dagger} U_i U_j) \right]\ ,
\label{ape}
\end{equation}
with $c$ and $f$ free smearing parameters.   Here ${\rm Proj_{SU(N)}}$ means the projection onto the SU(N) matrix.   In this paper, we took $n_{ape}=10$, $c=0.4$ and $f=0.081$.  

The correlation functions of smeared operators are given by 
\begin{equation}
C(t; \gamma_A^\ell, \gamma_B^{\ell'})=\frac{1}{2 \sqrt{N}}\sum_{n_0=0}^{2\sqrt{N}-1}{\rm e}^{-i \pi t n_0/(a\sqrt{N}) } {\rm Tr} \left[D_s^\ell \gamma_A D^{-1}(0)D_s^{\ell'} \gamma_B D^{-1}(p_0)\right].
\label{smeared_meson_prop}
\end{equation}
The computation of the traces and the inversion of
the Dirac operator is performed by means of a stochastic  evaluation based on the use of  $Z_4$ random sources~\cite{Eicker:1996gk,Gusken:1998wy}.
Let $z(\alpha,\beta,\gamma)$ be the source vector having color ($U_\mu$) index $1\le \alpha \le N$, spatial ($\Gamma_\mu$) index 
$1\le \beta \le N$ and Dirac ($\gamma_\mu$) index $1\le \gamma \le 4$. The source vectors take values
$z(\alpha,\beta,\gamma)=\frac{1}{\sqrt{2}}(\pm1 \pm i)$ and satisfy  
$\left< z^{*}(\alpha',\beta',\gamma')z(\alpha,\beta,\gamma)\right>=\delta_{\alpha'\alpha}\delta_{\beta'\beta}\delta_{\gamma'\gamma}$, when averaged over random sources.  Now,
we can replace the trace in Eq.~(\ref{smeared_meson_prop}) by the following expression:
\begin{eqnarray}
&&{\rm Tr} \left[[D_s^\ell \gamma_A D^{-1}(0)D_s^{\ell'}  \gamma_B D^{-1}(p_0)\right] \\
&=&\left<z^{\dagger} D_s^\ell \gamma_A D^{-1}(0)  D_s^{\ell'} \gamma_B D^{-1}(p_0)z\right> \\
&=&\left<z^{\dagger} D_s^\ell \gamma_A \gamma_5Q^{-1}(0)  D_s^{\ell'} \gamma_B \gamma_5 Q^{-1}(p_0)z\right> \label{average}
\end{eqnarray}
where $Q(p_0)=D(p_0)\gamma_5$.
To account for the inversion of the Dirac operator, we solve the following $2\sqrt{N}$ equations for $y(p_0)$
\begin{equation}
Q(p_0) y(p_0) = z,\ \ \ \  p_0 = \frac{\pi n_0}{a \sqrt{N}} \quad {\rm with} \,  0\le n_0 \le 2\sqrt{N}-1
\end{equation}
and the following equation for  $x^A_\ell$
\begin{equation}
Q(0)x^A_\ell= \gamma_5 \gamma_A^\dagger D_s^\ell z  \\
\end{equation}
Averaging over random sources, and taking into account that $D_s$ is a Hermitian operator, we have
\begin{equation}
\left<(x_\ell^A)^\dagger D_s^{\ell'} \gamma_B \gamma_5 y(p_0) \right>=\left<z^{\dagger} D_s^\ell \gamma_A \gamma_5Q^{-1}(0)  D_s^{\ell'} \gamma_B \gamma_5 Q^{-1}(p_0)z\right>\, .
\label{average2}
\end{equation}
In the actual average, we use 5 random sources. 
We are using the Conjugate Gradient inversion, so $Q^{-1}$ actually means $Q^{-1}=QQ^{-2}$.

For each value of $b$, many different values of
$\kappa$ have been used (the list will be given later). By fitting the
dependence of the PCAC quark mass (defined later) on $\kappa$, one can determine $\kappa_c(b,N)$, which is the chiral limit at which this mass vanishes. The list of values obtained in our simulation are
collected in Table~\ref{table_kc}.

As explained in the previous section,  we have  also studied twisted mass fermions with maximal twist~\cite{Frezzotti:2000nk}. This corresponds to a modified Dirac operator of the form 
\be
\label{DTM}
\DTM^{\pm}= \DWD\big|_{\kappa=\kappa_c} \pm i 2 \kappa_c \mu \gamma_5
\ee
where the Wilson-Dirac operator is computed at the value $\kappa_c(b,N)$ determined  earlier with Wilson fermions. After a chiral rotation of the fermion fields this can be written as the Dirac operator of a fermion field of mass $\mu$ plus an irrelevant modified Wilson term. The interesting  bilinear operators from which we compute correlators  involve two different fields having opposite values of the $\mu$-term.  It is equivalent to having non-singlet bilinear operators involving two flavours. This modified Dirac operator has many advantages. For example, the constant $\mu$ is proportional to the quark mass and monitors the explicit violation of chiral symmetry. It has also important advantages from the point of view of lattice artefacts~\cite{Frezzotti:2003ni} and most importantly for our purposes allows the determination of the pseudoscalar decay constant without unknown renormalization factors~\cite{DellaMorte:2001tu} (for a review of properties and results, plus a list of other relevant references we refer the reader to Ref.~\cite{Shindler:2007vp}). 

As shown in the last section,   the original formula for the correlator in momentum space Eq.~\eqref{corr_mom} involves a momentum sum. In principle choosing  different ranges only produces finite $N$ corrections. A priori it seems that the formula involving the momentum sum is harder to obtain since one would need to invert the propagator for all values of $q$. However, there is a trick that can be used that costs no extra effort. The idea is for each gauge field configuration to generate a random value of $q$ within the right range and with a uniform distribution. In particular if we consider that the quark field lives in an infinite lattice, for every gauge field configuration we can generate a random angle $\alpha_\mu$ and replace the link in the Dirac operator 
as follows:
\be
U_\mu \longrightarrow {\rm e}^{i\alpha_\mu} U_\mu
\ee
and then apply the same source and inversion procedure as explained earlier. The method produces results at no extra computational cost and we verified that for the free theory the method gives the right propagator. Since the modified links are in U(N) rather than SU(N), we did not impose the unit determinant condition in the Ape smearing procedure. 

For Wilson fermions we used both methods giving compatible results within errors. For consistency, all the data and masses presented here for the Wilson fermions were obtained with  the simple version with no momentum sum. However,  our twisted mass results are obtained with the random angle method. Comparison of the results obtained with both methods provides an explicit test that finite $N$ corrections are under control.

In relation with the previous comment, it is clear that stronger finite $N$ corrections are expected when we approach the   chiral limit, since the 
effective lattice size  becomes small in units of the inverse pion mass. To avoid this problem,  all our results have been obtained for large enough values of
$m_\pi l$. The results are then extrapolated to the chiral limit. We  should point out that because of our large values of $N$ this has less problems that for small ones, since chiral logs are strongly
suppressed. This is one advantage of using our method.

Finally, as explained earlier, once we obtain the expectation values of
the correlators as a function of momenta we Fourier transform them to
a time dependent function. For large enough time-separations these
correlation functions would decay exponentially in time, the coefficient of
which determines the minimal mass in that channel in lattice units. 
This is in short the procedure. In practice this becomes much more
complicated and requires an efficient treatment of all data. This will
be explained in the following subsection.

%% file: mass_computation.tex
\label{s:mass_comp}
As just mentioned, meson masses are extracted from the exponential decay in time of appropriate correlation functions.
Although such correlators receive contribution from a tower of states, 
the lowest mass in the corresponding channel dominates at large times and can be determined from the coefficient of the late single-exponential decay.
Nonetheless, this is not as straightforward as it may seem; the determination of the large time decay in an actual simulation faces two problems: the fact that the time extent of the lattice is finite and  
the deterioration of the signal to noise ratio at large time separation. Therefore, 
for a precise determination of the ground-state mass it is advantageous to work with operators for which the
projection onto the ground state is maximized and the single-exponential decay sets in early on.  
Smearing is one step in that direction; it increases the overlap with the ground state wave function by extending 
the operator range at distances comparable with the inverse wavelength of the particle. 
Beyond that, further improvement can be obtained by using a variational approach~\cite{\GEVPONE,\GEVPTWO}.
The idea is very simple and is based on constructing a linear combination of operators with the same
quantum numbers, with coefficients tuned to optimize the ground state projection. 
We will describe below the particular implementation of the variational procedure used in this work.

To start with, consider a basis of $N_{op}$ operators with the right quantum numbers, in terms of which a $N_{op} \times N_{op}$ correlator matrix is built:
\be
C_{a b} (t) = \langle O_a^\dagger(t) O_b(0) \rangle .
\label{eq:c_orig}
\ee
The variational strategy looks for solutions to the generalized eigenvalue problem (GEVP):
 \be
  C_{ab} (t_1) v_b^{(i)}(t_1,t_0)  = \lambda^{(i)}(t_1,t_0) \,   C_{ab} (t_0) v_b^{(i)}(t_1,t_0)  ,            
\ee
for given $t_1$ and $t_0$, with $t_1>t_0$. This is obviously equivalent to solving the standard eigenvalue problem for the matrix:
\be
C^{-\half} (t_0) C (t_1) C^{-\half} (t_0),
\ee
with eigenvalues $\lambda^{(i)}(t_1,t_0)$ and corresponding eigenvectors $C^{1/2}(t_0) v^{(i)}(t_1,t_0)$.
The correlation matrix given in eq.~\eqref{eq:c_orig} can be rotated to the $v^{(i)}$ basis, leading to the matrix:
\be
\tilde C_{i j}(t, t_1,t_0) = v^{(i)}_a(t_1,t_0)  C_{ab} (t) v^{(j)}_b(t_1,t_0),
\label{eq_c_opt}
\ee
which is diagonal at $t=t_1$ with eigenvalues given by $\lambda^{(i)}(t_1,t_0)$.
With a complete basis of operators and for infinite time extent, one would obtain $\lambda^{(i)}(t_1,t_0) = \exp(- m_i (t_1-t_0))$.
The objective of the procedure is to select the set of operators and the choice of $t_0$ and $t_1$ to optimize the projection onto
the lowest states, in particular the ground state. 
The standard GEVP proceeds by repeating the diagonalization at all values of $t=t_1$, looking for a plateaux in the effective masses extracted from
$\lambda^{(i)}(t_1,t_0)$. Given the limited time extent of our lattice, we have preferred instead to fix the value of $t_1$
and work with the correlator matrix defined by $\tilde C(t,t_1,t_0)$. To determine the ground state mass in each channel, we use the diagonal matrix element
related to the largest eigenvalue $\lambda_{\rm max}(t_1,t_0) = \max_{i} \lambda^{(i)}(t_1,t_0) $. 
Let us denote by $v^{\rm max}(t_1,t_0)$ the corresponding eigenvector in terms of which we introduce an {\it optimal}, within the given choice of operators and selected 
values of $t_0$ and $t_1$, correlation function:  
\be
C^{\rm opt}(t, t_1,t_0) = v^{{\rm max}*}_a(t_1,t_0)  C_{ab} (t) v^{\rm max}_b(t_1,t_0).
\ee
The ground-state mass is extracted in the usual way from the exponential decay at large times of this correlator.

\begin{figure}[t]
\centering
\includegraphics[width=0.80\linewidth]{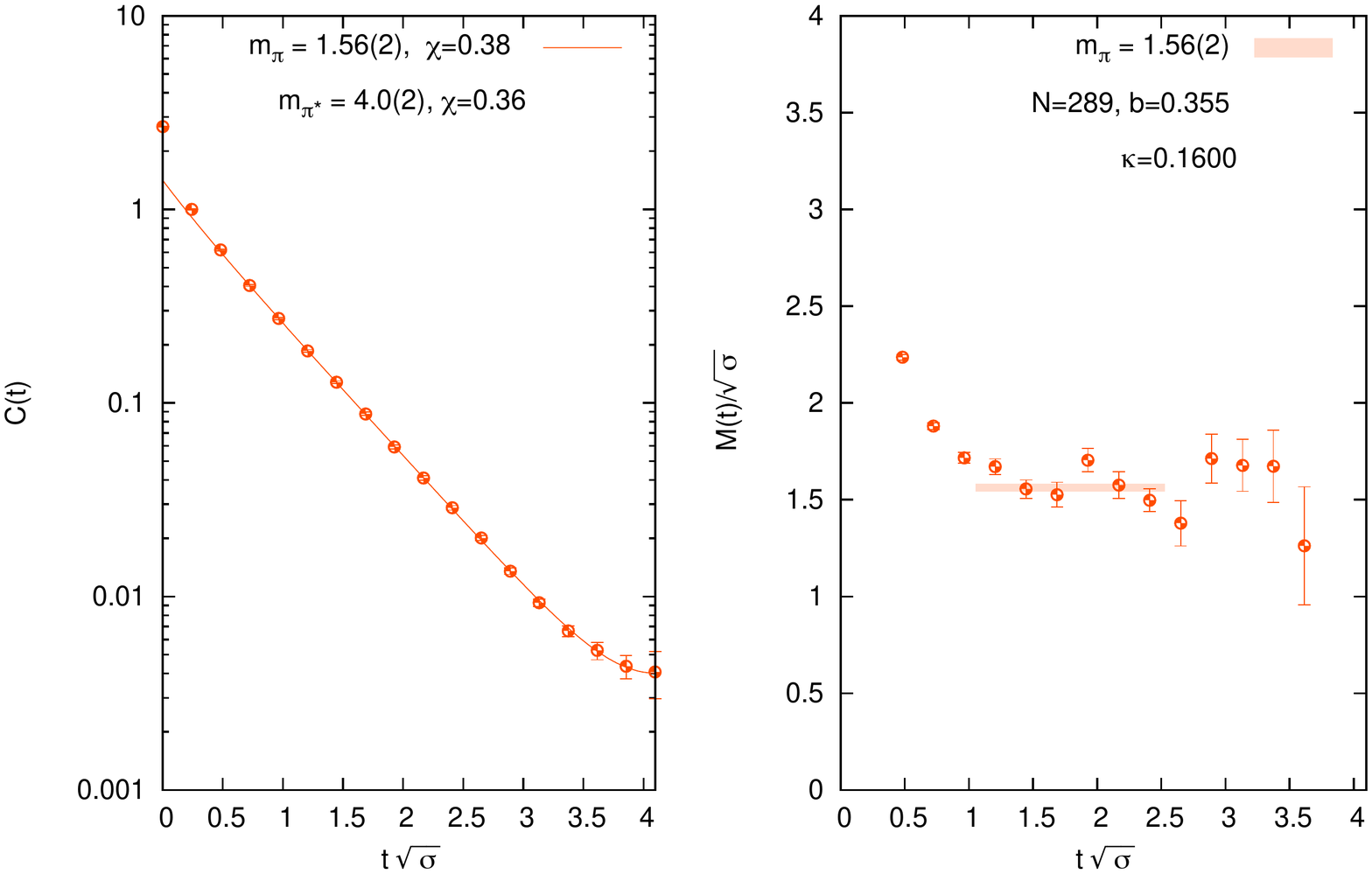}\\
\includegraphics[width=0.80\linewidth]{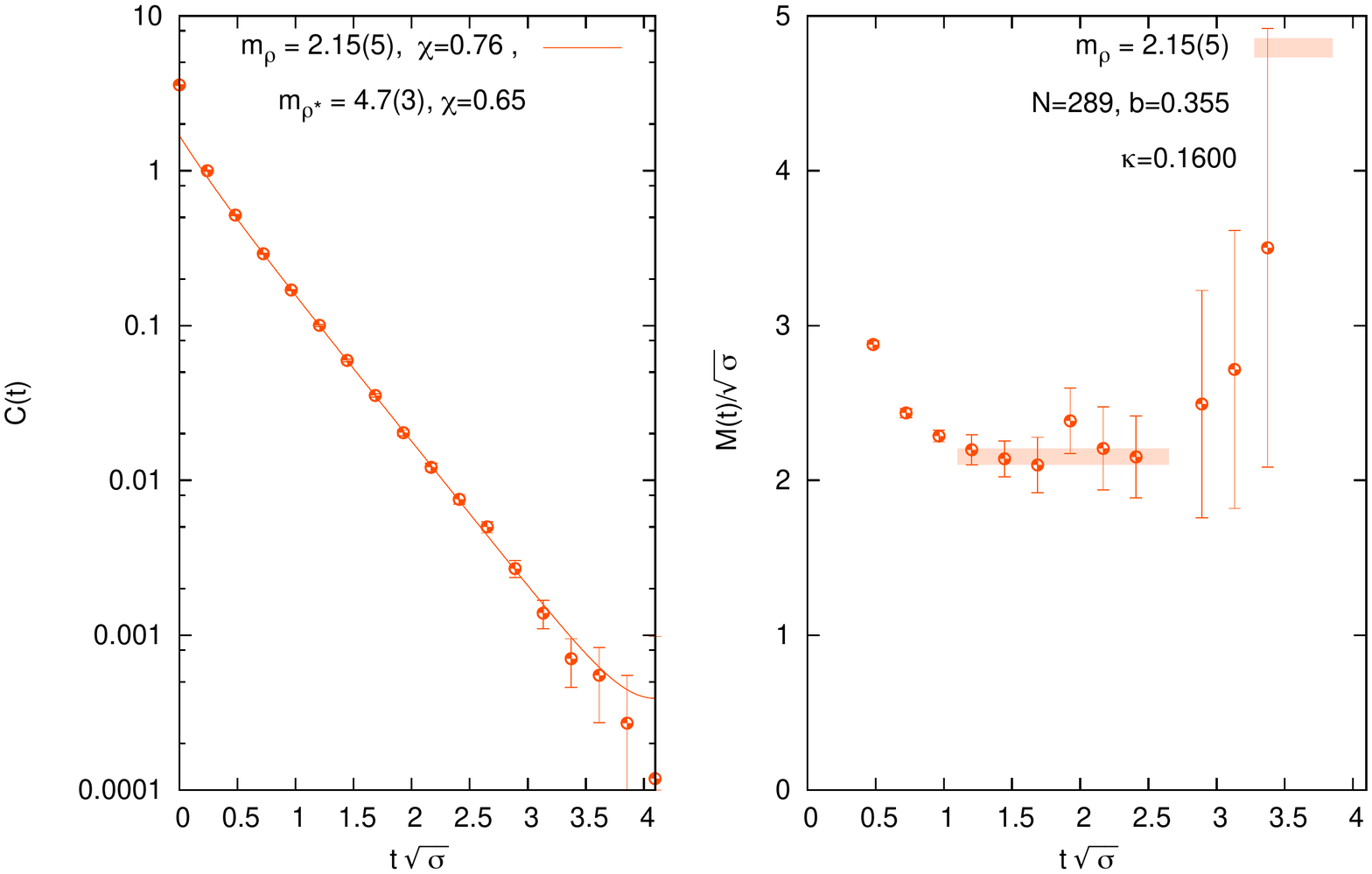}
\caption{Plots on the left show the time dependence of the {\it optimal} correlator for pseudoscalar (top) and vector (bottom) mesons.
Plots on the right show the time dependence of the effective masses. The red band indicates the mass of the ground-state
determined from the fit to the correlator with the length of the band indicating the fitting range and the width indicating the jacknife error.
}
\label{f_ex1}
\end{figure}
\begin{figure}[t]
\centering
\includegraphics[width=0.70\linewidth]{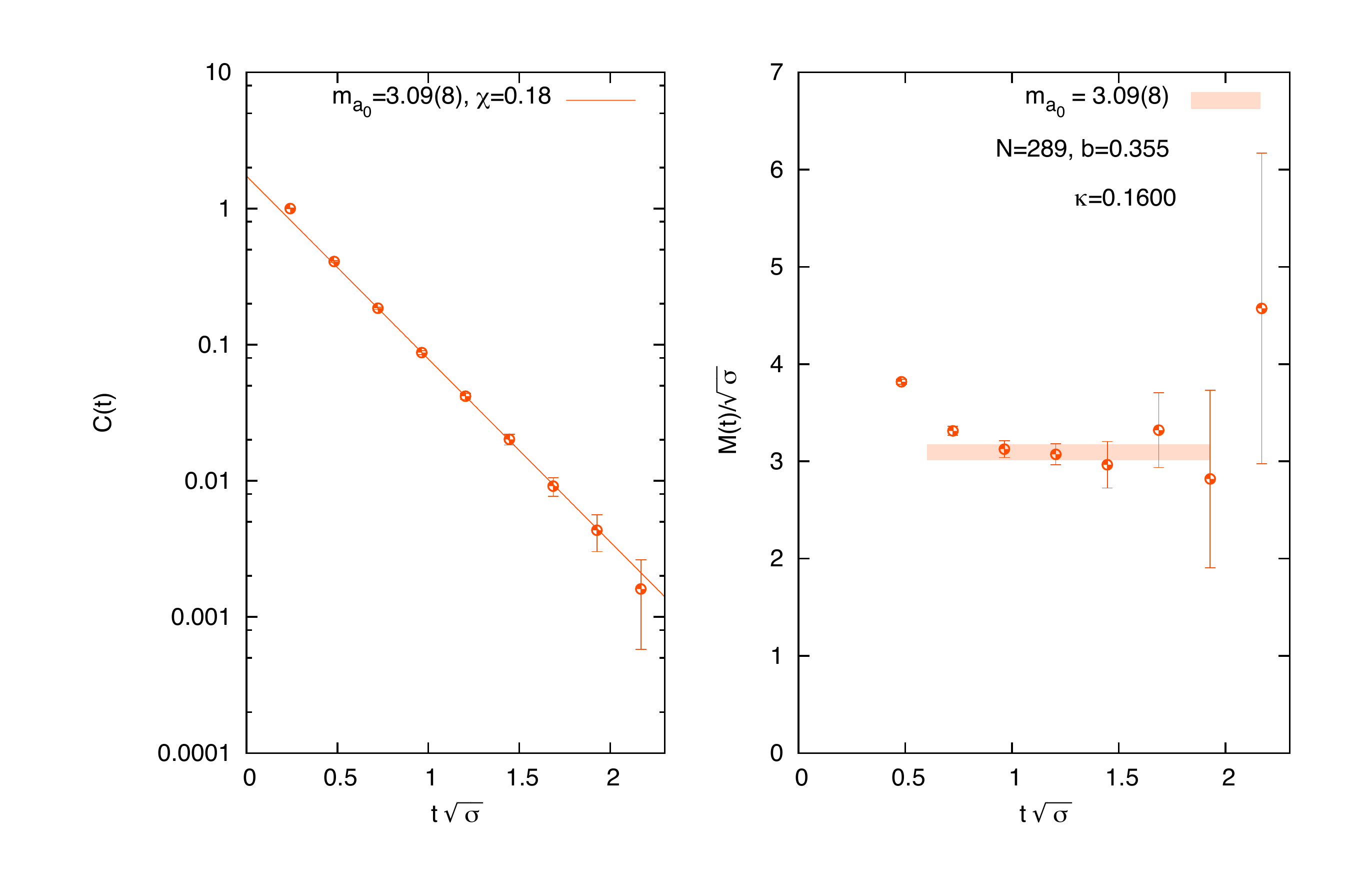}\\
\includegraphics[width=0.70\linewidth]{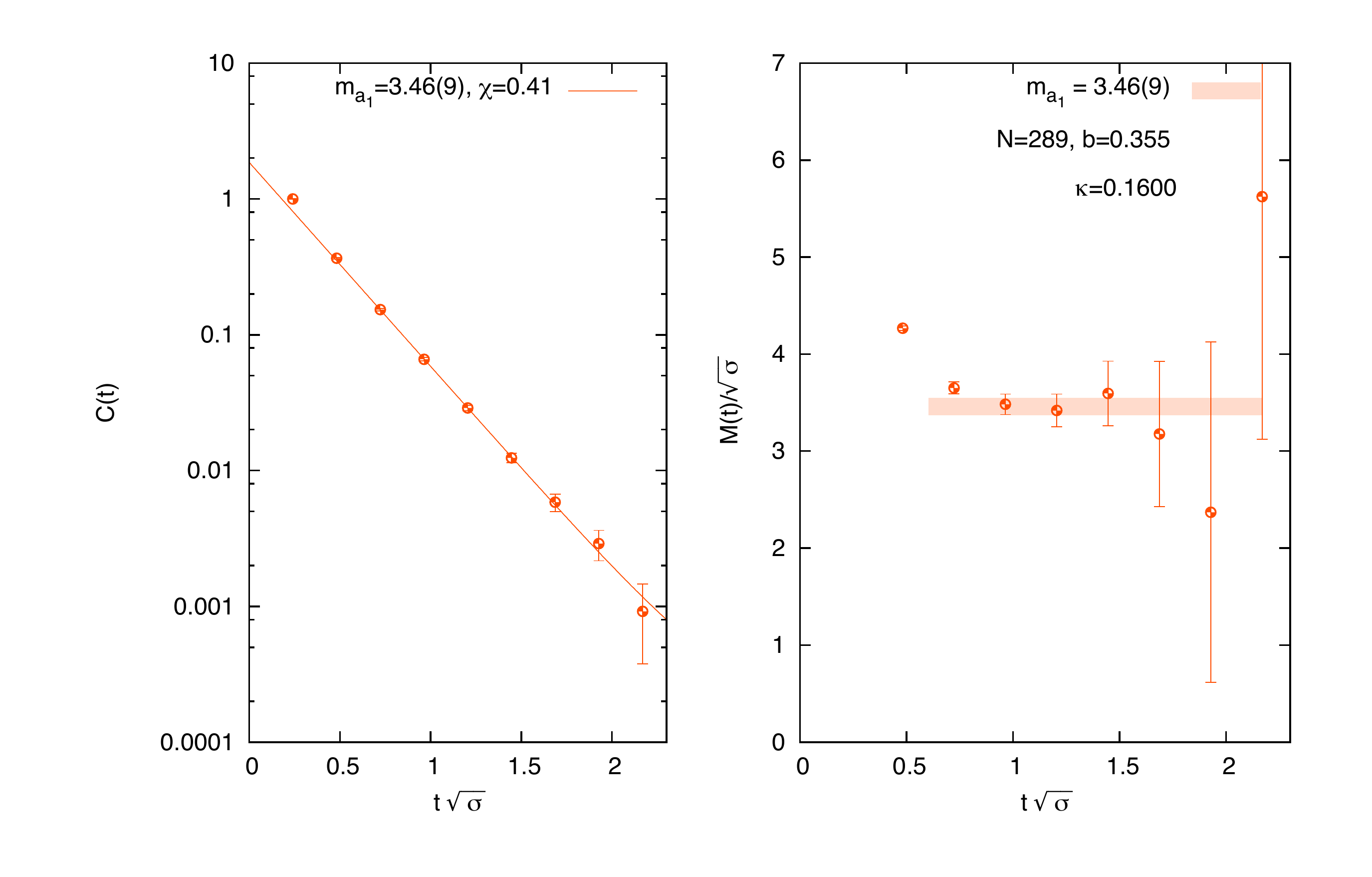}\\
\includegraphics[width=0.70\linewidth]{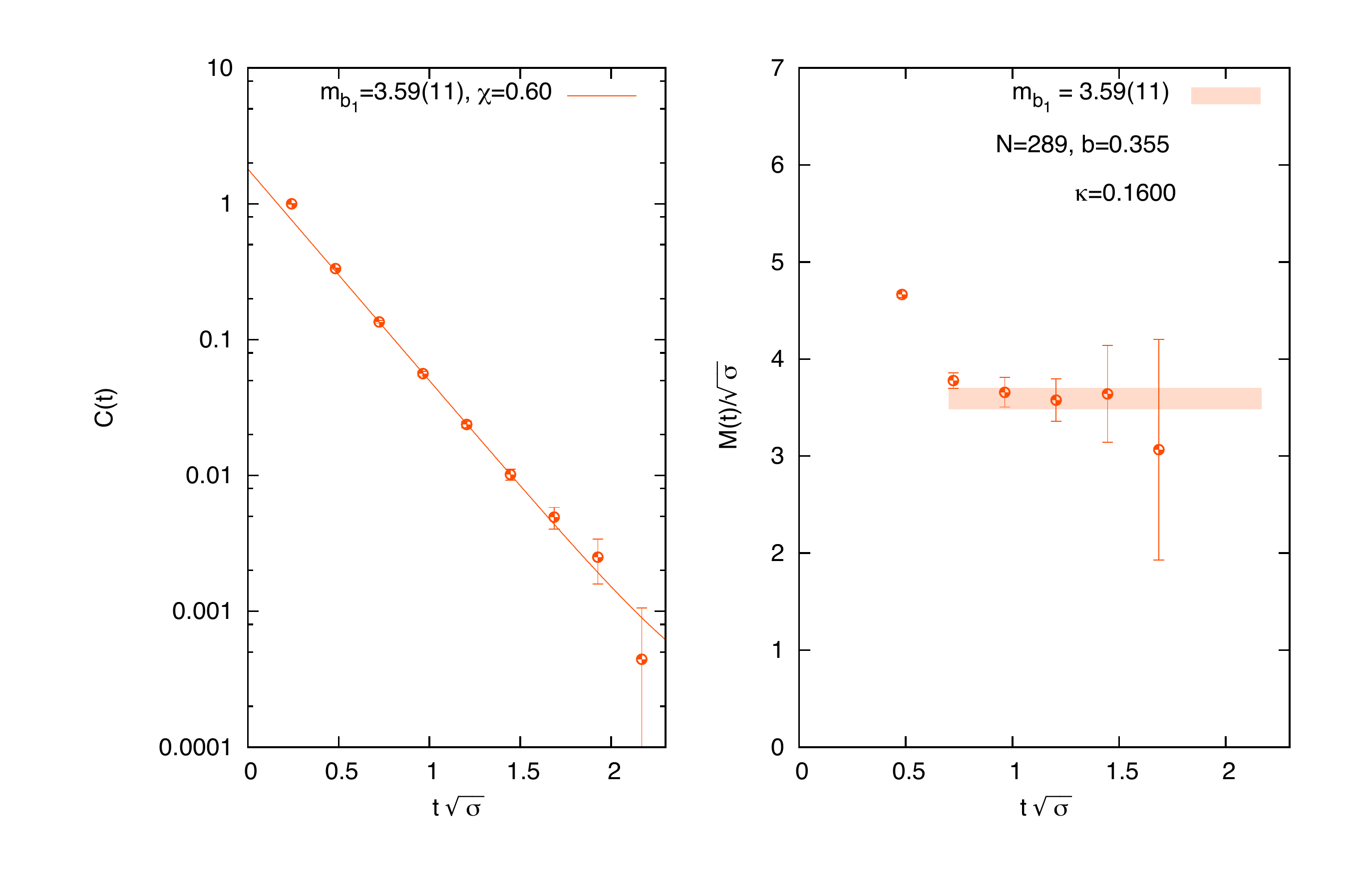}
\caption{As in fig.~\ref{f_ex1} for, from top to bottom, $a_0$, $a_1$ and $b_1$ mesons.
}
\label{f_ex2}
\end{figure}

For the specific implementation in this work, we have taken $t_0=a$ and $t_1=2a$, with $a$ the lattice spacing. The choice obeys to a compromise to maximize
the projection onto the ground state without affecting the signal to noise ratio which becomes worse for larger values of $t_i$. We have tested that other choices in the
range of values of $t_i$ below $4a$ give compatible results.
The operators in the basis are constructed by using different smearing levels of the quark bilinear operator as described in section~\ref{s:observables}.
We have considered up to 10 operators corresponding to fermion smearing levels: 0, 1, 2, 3, 4, 5, 10, 20, 50, 100.
Ground-state meson masses in each channel are extracted by fitting the {\it optimal} correlator to a hyperbolic-cosine. The optimization procedure involves, first, a choice of the operators in the basis and, second,  the choice of fitting range $[t_{\rm min},t_{\rm max}]$. These choices are the main sources of systematic errors in the mass determination. We will discuss the selection criteria in sec.~\ref{s:errors}. Let us just mention here that they respond to two general considerations:  the reduction of excited state contamination and the limitation of finite size effects and signal reduction at large times. While smearing improves the projection onto the ground state, at the same time it leads to larger finite size effects and larger statistical errors at large time, for this reason we vary the selection of operators in the basis to limit the maximal amount of smearing. In general we have used the full basis in the pseudoscalar channel and up to 50 smearing levels for the other meson channels. 

In order to illustrate the quality of the fits that can be attained with this procedure, we present in figs.~\ref{f_ex1}, ~\ref{f_ex2}  one example in each meson channel. 
The figure corresponds to a lattice with $N=289$ at $b=0.355$ and $\kappa=0.16$.
For each channel, we display the time dependence of the optimal correlator (left) and the effective mass (right). 
The red band in the plot shows, for illustration, how the determination extracted  from the hyperbolic cosine fit compares with the effective mass plateaux, 
 with the length of the band indicating in each case the fitting range $[t_{\rm min},t_{\rm max}]$.

In addition to the ground state mass, we have also attempted to estimate the mass of the first excited state in each meson channel. 
For that purpose, we have performed double exponential fits keeping 
the ground state mass, $m$, fixed to the one extracted from the optimal correlator. To be specific, we look at correlators derived including in the variational basis the 4 lowest smearing levels. This correlator is fitted, in the range $t\in[(2\sqrt{\sigma})^{-1},t_{\rm max}+a]$, with a combination of 
two hyperbolic cosine functions with one mass parameter free and the other fixed to the ground-state mass. 
This allows to determine the first excited state mass $m^*$. 
The correlator fits for the pseudoscalar and vector mesons shown in fig.~\ref{f_ex1} include these two contributions with $m$ and $m^*$ fixed to the values determined in this way.

%% file: errors.tex
The goal of this paper is to provide values for the mesons masses and
decay constants for the large $N$ gauge theory based on first
principles that can be used as a test for other estimates based on  other 
types of ideas and approximations. For that purpose the final numbers have
to be supplied with trustworthy errors. Some of these errors are
statistical. These are indeed the simplest conceptually. In this paper
we have employed the well-known jacknife method, based on splitting
the sample into several groups of equal size, computing the quantity
in question by averaging over  all but one of these groups, and
equating the error to the dispersion of these values.   

A much more difficult task is the estimate of systematic errors. These
have different sources and are not necessarily symmetric around the
mean value. Most of the errors that we will be concerned with are typical
of all lattice calculations, but our methodology enhances or reduces
some of those. A very important characteristic of the lattice approach is that
all the errors can be estimated.  

\subsubsection{Simulation and inversion errors}
Here we gather together the errors that are characteristic of the
numerical procedure. We are fairly confident that the generation of
configurations is safe. We have previously carried the same procedure
for much larger values of $N$ (up to 1369) and much larger values of
$b$ (up to 0.385) and found no relevant systematic effects. Thus, this
should even be more so in the more restricted range of parameters that
we are exploring here. Furthermore we also played with different
thermalization steps and updates per measurement and found no
deviations. 

Concerning the methodological aspects entering in the evaluation of the
observables, we made several tests
to quantify the errors committed. In particular we used the same
programs and methods for the free field theory (coupling
$b\longrightarrow \infty$). The advantage is that the propagators and
correlators can be computed analytically in this case, so that the
comparison with data gives an idea of the numerical errors involved.
In this respect we tested the effect of taking a given number of
sources in the calculation of the meson correlators. With 200 sources
we found complete agreement between the analytic and numerical
computation within errors. With 5 sources some systematic deviations
are observed for small values of the quark mass. However, the main
effect turns out to be in the meson correlators for $p_0=0$ which
produces the addition of a small  constant  to the correlator. In
configuration space the effect only shows up at large correlation times. 
Thus, in our mass extraction algorithms we have set the upper limit of the fitting range  
to avoid this region, which is also the most affected by other statistical and
systematic errors.

In estimating the final meson masses and decay constants systematic 
errors are expected to arise from finite lattice spacing and finite
size effects. In our particular construction, finite size is traded
by finite $N$. There is  an  antagonistic interplay among these two 
error sources. For discretization errors to be small, $a(b)$ should be
small,  while for finite size errors to be small $\sqrt{N} a(b)$
should be large. This is pretty tight given that most of our data are 
obtained for $\hL= \sqrt{N}=17$. 

The sensitivity to finite size effects is larger when there are large
correlation lengths involved. Thus, in our measurements we have stayed 
away from the worse region, keeping the pion mass times the effective
length large enough. Anyhow,  to monitor and estimate the size of 
these errors we have studied the most sensitive quantities with  two
other values of $N$ (in addition to  our four  values of $b$). 

It is important to realize that the finite-size effects in the meson
correlators have two sources. On one side, the gauge field
configurations generated with the TEK model correspond to a finite
volume of size $\hL^4$, as Wilson loop expectation values show. On the
other hand the  quarks can be made to propagate in a box of the same
size or larger. These two options for the quark box size can be  easily 
implemented, as explained in the previous section. Thus, the effect can be monitored.
We tested this first for the case of free fermion fields  ($b\longrightarrow \infty$) 
in which  the meson propagator can be computed
analytically (by a momentum sum) as well. Indeed, the effects of using
different quark box sizes are sizably smaller for the twisted mass Dirac operator
than for the Wilson-Dirac one.

For our dynamical configurations at  finite $b$  we have also monitored the effect of 
the fermion propagation volume. The results  obtained with the simplest formula,
which involves no explicit sum over spatial momenta, corresponds to a fermion box of size
$\hL^4$. The results corresponding to an infinite quark box were obtained, as explained before, 
by generating momenta $p$ randomly with a uniform  distribution for each configuration.
The procedure does not involve any additional cost in computer time
and is seen to give the correct results in our free fermion case. We
did not find significant differences within errors between  the random
momenta results and the others.  

\begin{figure}[ht]
\centering
\includegraphics[width=0.90\linewidth]{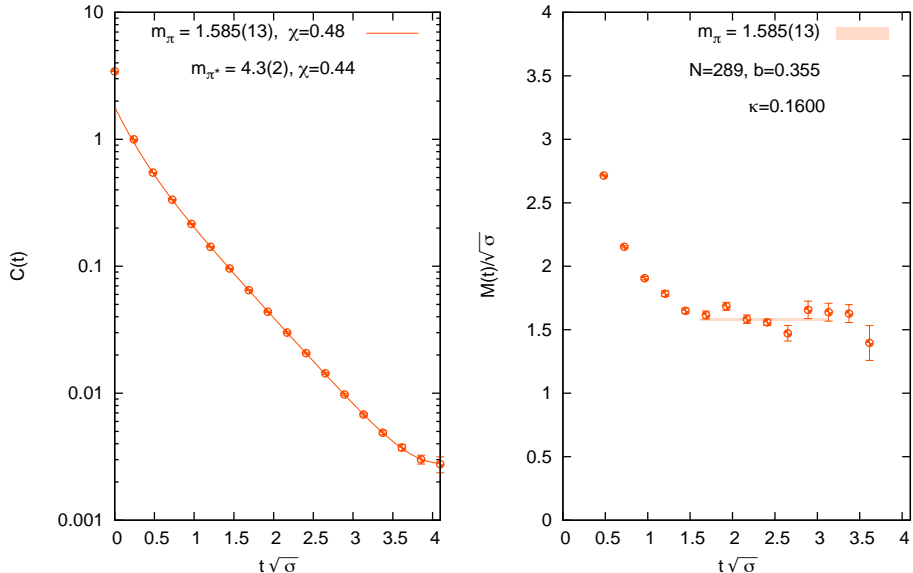}
\caption{The plot on the left shows the time dependence of the  correlator in the pseudoscalar channel obtained from the variational analysis including operators up to smearing level 10. The plot on the right shows the time dependence of the effective masses. The red band indicates the mass of the ground-state
pion determined from the fit to the correlator with the length of the band indicating the fitting range and the width indicating the jacknife error.
}
\label{fig_op7}
\end{figure}

Finally, we come to the systematic errors which arise from the
procedure to extract the masses from the time-dependent meson  correlators.
This is by far the largest source of systematic errors. In our
procedure masses are obtained from a fit to an exponential decay of the
time-correlation function of a properly selected combination of mesonic
operators with the appropriate quantum numbers. The results then depend
on the selection of the operator, the region of fitting and the
functional form chosen for the fit.  As discussed in sec.~\ref{s:mass_comp} , the mesonic operator in each channel is extracted from a variational analysis aimed at improving the projection onto the ground state. The basis for this variational analysis is constructed in terms of smeared operators with smearing level 0, 1, 2, 3, 4, 5, 10, 20, 50 and 100.  The results that will be presented correspond to a particular choice that we consider {\it optimal} and includes all operators up to smearing level 20, 50 or 100, depending on
the state under study and the physical volume of the lattice. The selection of fitting range, $[\tmin,\tmax]$ responds to two general considerations: $\tmin$ 
is tuned to reduce contamination from excited states, and $\tmax$ to limit finite size effects and the poor signal to noise ratio at large times.  Concrete choices for these quantities depend not only on the channel considered but also on the amount of smearing of the operators in the basis;  including operators with larger smearing level improves the projection onto the ground state and allows to reduce the value of $\tmin$, but, at the same time, the signal to noise ratio deteriorates at large times, limiting the value of $\tmax$.  
As a rule of thumb, given a meson state and operator choice, $\tmin$ is fixed in physical units, around the value where the effective mass plateaux sets in, and $\tmax$ is set to a fraction of the maximal time separation ($a\sqrt{N}$). 
In order to estimate the systematic effects in the selection of smearing and fitting ranges, we have analyzed the spectrum obtained if only operators with smearing level below 10 are included in the variational basis; as an example we display in fig.~\ref{fig_op7} the fits to the pseudoscalar correlator, corresponding to the case presented previously in fig.~\ref{f_ex1}. The masses obtained are in good agreement giving $m_{\pi}/ \sqrt{\sigma}= 1.563(21)$ and 1.585(13). In the general case, the results of this exercise lead to masses that are compatible within statistical errors
to the ones derived with the optimal operator.

Finally, concerning the functional form used for the fit, the ground state masses are derived from a single hyperbolic cosine fit performed in the range
$[\tmin,\tmax]$, however, in the case of the heaviest states ($a_0$, $a_1$ and $b_1$) we also included a constant contribution that improved the $\chi^2$ of the fits at large times. In most cases the constant turned out to be small and the resulting masses compatible within errors with those obtained with no constant added.

%% file: pion_channel.tex
\subsubsection{Chiral behaviour and the pion mass}

\begin{figure}[t]
\centering
\includegraphics[width=0.49\linewidth]{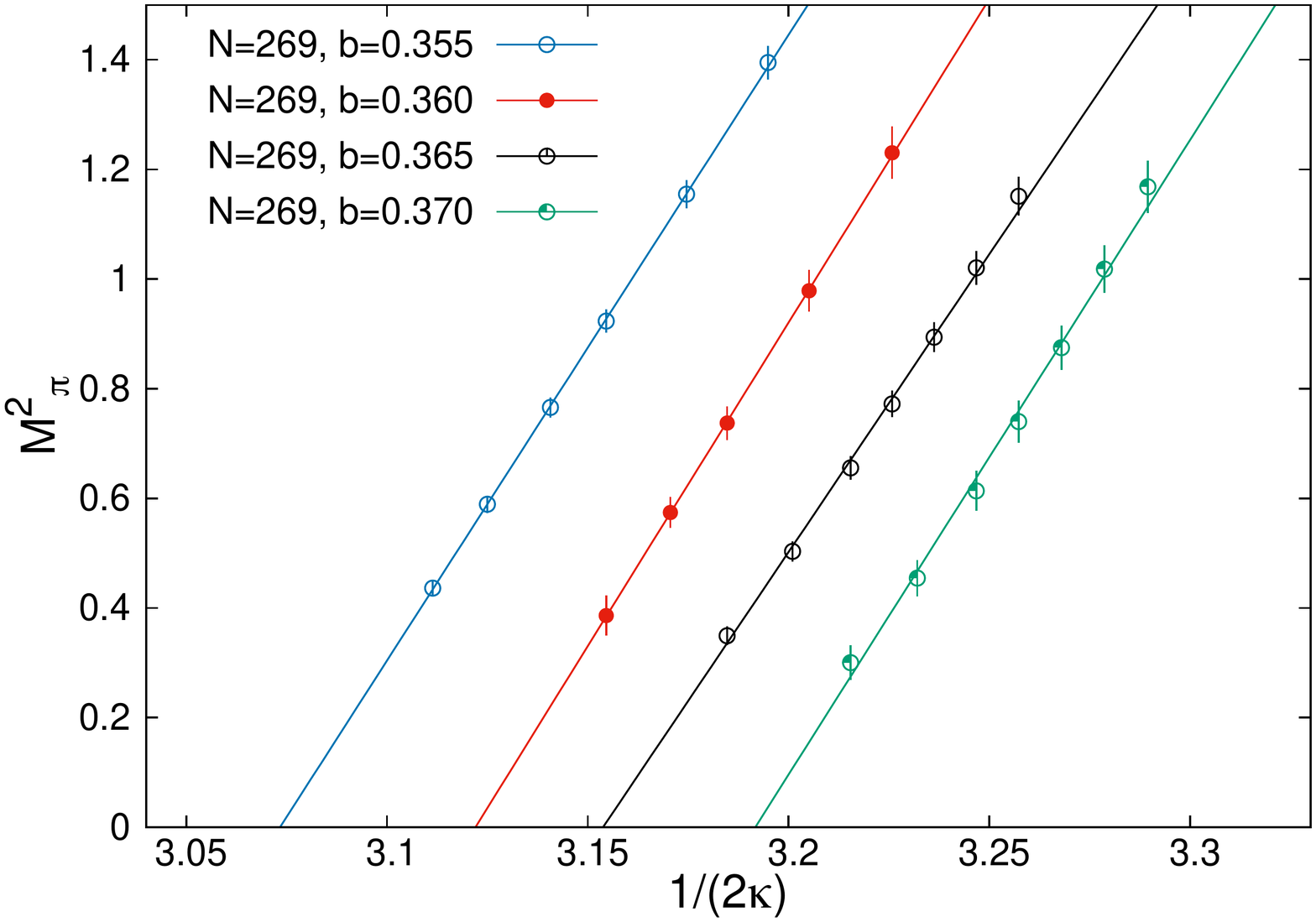}
\includegraphics[width=0.49\linewidth]{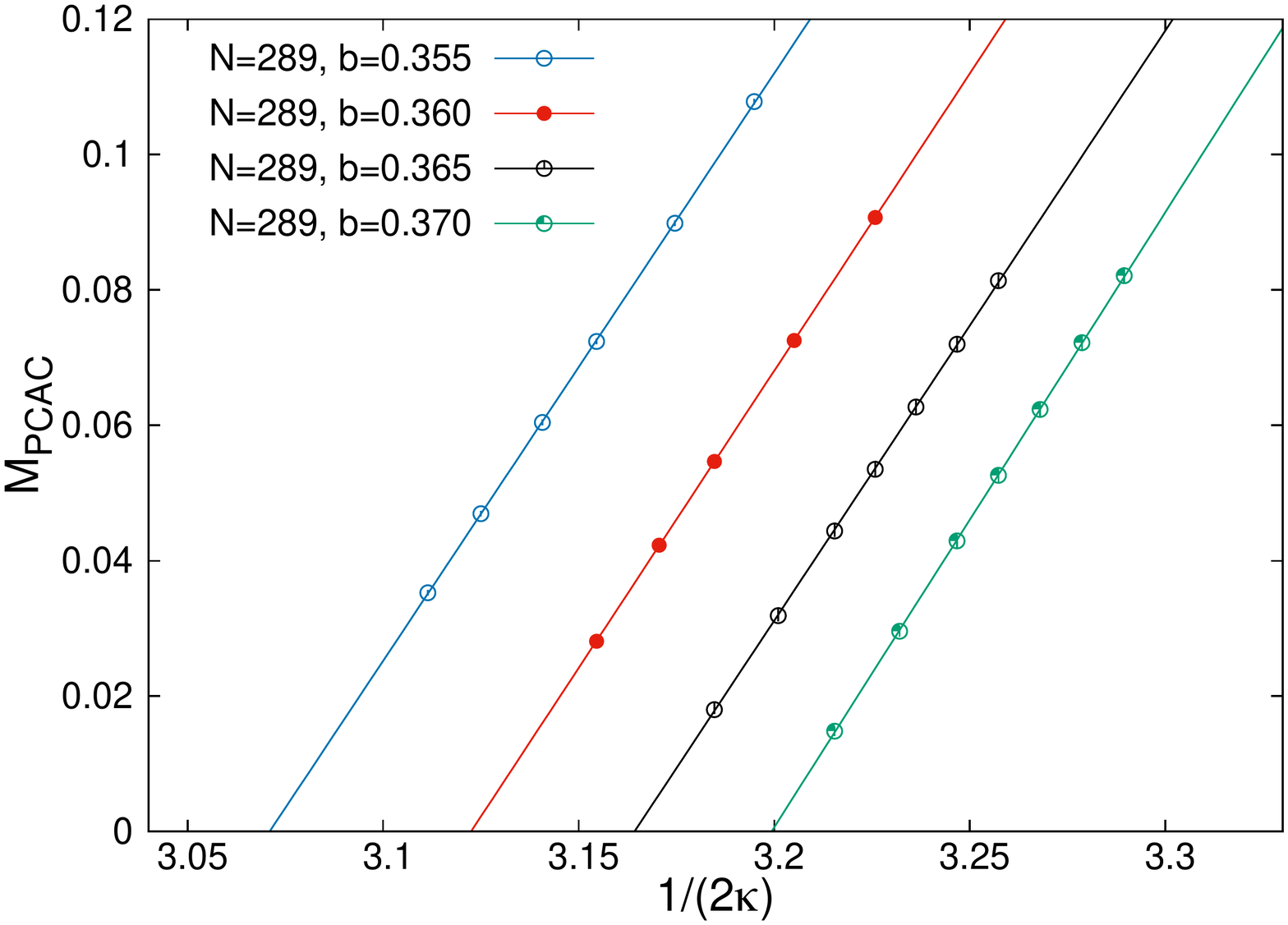}
\caption{Dependence of the pion mass square (left) and the PCAC mass (right) in units of the string tension as a function of $1/(2 \kappa)$.
}
\label{fig2}
\end{figure}

We start with the most important channel: that involving correlators
among pseudoscalar meson operators. Because of $\gamma_5$-hermiticity
the correlation function is positive definite. Our results show a very
neat exponential fall-off of the propagator for large enough
separations. Using the optimal operator, as explained in the previous
section, the exponential behaviour  extends to rather low separations.
Fig.~\ref{f_ex1} shows some characteristic correlators illustrating the
typical error size.   
From these correlators a fit allows us to
determine the mass of the lowest energy state, which we will call the pion. 
We show in Fig.~\ref{fig2} an example of the dependence of the square of this mass in
units of the string tension as a function of $1/(2 \kappa)$. We see that the
dependence is linear. Indeed, this is what we expect   from the
pseudo-Goldstone boson nature of the pion.   The quantity $1/(2 \kappa)$ is a
linear function of the quark mass. The vanishing of the pion mass
occurs when the quark mass vanishes and we recover the explicit chiral
symmetry, which remains spontaneously broken. In Table~\ref{table_kc} we
list $\kappa_c^\pi$,  the value of $\kappa$ at which our linear fit
predicts the vanishing of the pion mass. We also give the  slope and 
the square root of the Chi-square per degree of freedom of the linear fit
 ($\bar \chi \equiv \sqrt{\chi^2/\#dof}$). The slope  is
scaled by $a(b)$ in string tension units. This scaling is the expected
one if both the lattice pion mass $M_\pi = m_\pi a(b)$ and the bare
quark mass $M_q\equiv \frac{1}{2 \kappa}- \frac{1}{2 \kappa_c}=Z_S m_q a(b)$, are scaled with the lattice spacing as shown. 

\begin{table}[ht]
  \begin{center}
    \begin{tabular}{|c|c|c|c|c|c|c|c|}
\hline
$N$ & $b$     &  Slope$/\sqrt{\sigma}$ & $\kappa_c^\pi$ & $\bar \chi$&   $Z_P/(Z_A Z_S)$ & $\kappa_c$ & $\bar\chi$\\
\hline
169 & 0.355 & 10.4(1.4)  & 0.16247 (38) & 0.20 & 0.894 (47) & 0.16248 (15) & 0.08\\
169 & 0.360 & 11.2(2.5)  & 0.15983 (47) & 0.26 & 0.0.917 (37) & 0.159827(91) & 0.50\\
\hline
289 & 0.355 & 11.4(3)  & 0.162689(96) & 0.19& 0.8681(62) & 0.162815(31) & 0.44 \\
289 & 0.360 & 11.8(7)  & 0.16015 (20) & 0.16 & 0.878 (15) & 0.160129(57) & 0.11\\
289 & 0.365 & 10.9(4)  & 0.15854 (11) & 0.66 & 0.872 (14) & 0.158014(45) & 0.22\\
289 & 0.370 & 11.6(6)  & 0.15666 (16) & 0.66 & 0.908 (15) & 0.156283(43) & 0.18\\
\hline
361 & 0.355 & 11.2(7)  & 0.16287(17)  & 0.06 & 0.840 (22) & 0.162978(82) & 0.07\\
361 & 0.360 & 11.1(6)  & 0.16022(15)  & 0.15& 0.873 (17) & 0.160243(49) & 0.01 \\
361 & 0.365 & 12.0(1.2)& 0.15815(22)  & 0.19 & 0.917 (44) & 0.158008(99) & 0.28\\
361 & 0.370 & 11.1(9)& 0.15681(18)  & 0.10& 0.884 (26) & 0.156380(49) & 0.13 \\
\hline
    \end{tabular}
  \caption{Determination  of the critical $\kappa$ from the linear dependence of the pion mass square ($\kappa_c^\pi$) or the PCAC mass ($\kappa_c$) on $1/(2\kappa)$. We also provide the slopes of the linear fits. In the case of the PCAC mass, the slope determines the ratio of renormalization constants $Z_P/(Z_A Z_S)$. In the case of the pion mass square we express the slope in string tension units. The quantity $\bar \chi$ denotes the square root of the reduced Chi-square
  ($\bar \chi \equiv \sqrt{\chi^2/\#dof}$). }
    \label{table_kc}
  \end{center}
\end{table}

An alternative  way to determine the onset of chiral symmetry is to
look directly at the Ward identity and the PCAC relation.  This allows 
the definition of a quark mass, called
PCAC mass $\MP$, whose vanishing signals the point where
the explicit breaking vanishes.
This involves expectation values of the axial vector and pseudoscalar 
operator between the vacuum and the pion at rest. Hence, the PCAC
mass is affected by operator renormalization and depends on the
operator that we use. On the lattice we  consider the ultralocal version
of the axial vector current and of the pseudoscalar operator. To compute it we follow a similar procedure as in Ref.~\cite{\BALIMESONS} by computing the correlation functions of the two operators in question with the optimized version of the
pseudoscalar operator used for the extraction of the pion mass. Hence, the values of the $\MP$ masses are determined by fitting to a constant the ratio 
\be
\MP(t) = \frac{ C(t+a;\gamma_0\gamma_5, \gamma_5^{\rm opt}) -  C (t-a; \gamma_0\gamma_5, \gamma_5^{\rm opt})}{4 C(t;\gamma_5, \gamma_5^{\rm opt})}
\ee
This new mass is proportional to the quark mass $M_q$ with proportionality constant given by the ratio of renormalization factors $Z_P/(Z_A Z_S)$. Therefore, from the vanishing of $\MP$ one can extract an alternative derivation of the value of the critical $\kappa$, that should coincide in the continuum, infinite volume limit, with the one determined from the vanishing of the pion mass. The values of $\kappa_c$ extracted in this way, as well as the slopes and values of $\bar \chi$ are also given in Table~\ref{table_kc} -- an example of the quality of the fits is shown in Fig.~\ref{fig2}. This determination of $\kappa_c$ is more precise and will be taken as our best estimate of the chiral limit.
\begin{figure}[t]
\centering
\includegraphics[width=0.75\linewidth]{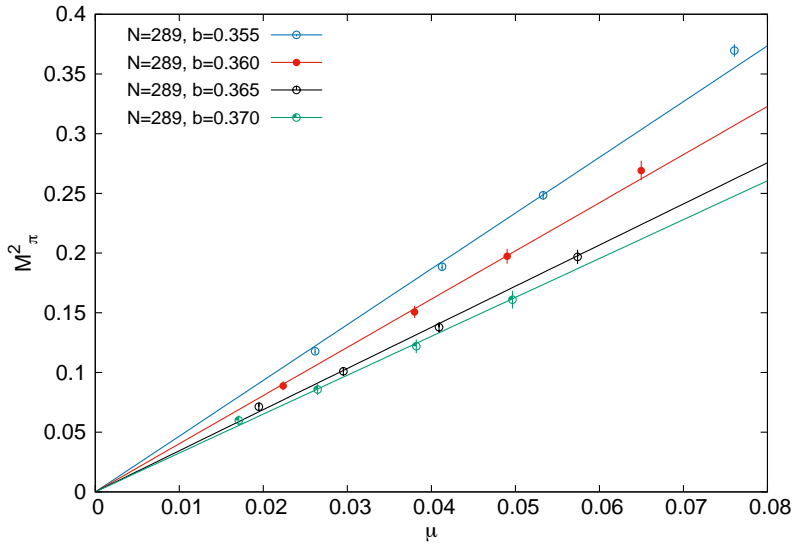}
\caption{Dependence of the pion mass square in lattice units as a function of the twisted mass parameter $\mu$ for $N=289$.}
\label{fig3}
\end{figure}

All the previous results concerned Wilson fermions. However, we have
also studied maximally twisted mass fermions~\cite{Frezzotti:2000nk}. For that one has to start with the
Dirac operator for massless Wilson fermions, hence at $\kappa= \kappa_c$. Then, as explained in  Eq.~\eqref{DTM},  one adds a mass term  $ \pm i  2 \kappa_c \mu  \gamma_5 $, where the opposite
sign is used for the two propagators involved in a meson correlator (see Eq.~\eqref{DTM} for the lattice twisted mass Dirac operator).
This, after a chiral rotation,  is equivalent to having two flavours of equal mass $\mu$  and computing the correlation function of non-singlet meson operators. The unfamiliar reader is invited to consult the literature~\cite{Frezzotti:2000nk,Frezzotti:2003ni,DellaMorte:2001tu,Shindler:2007vp}. By construction,  the pseudoscalar meson mass vanishes (approximately) at $\mu=0$. Furthermore, since $\mu$ is proportional to the quark mass,  the pion mass square should depend linearly on  $\mu$.   
Our results agree with this linear dependence  with intercept equal to zero. The fitted slopes  are given  in Table~\ref{table4}, again scaled
with the lattice spacing in string tension units. As an example, fig.~\ref{fig3} displays the data and the linear (one-parameter) fit for $N=289$.

\begin{table}[ht]
  \begin{center}
    \begin{tabular}{|c|c|c|c|c|}
\hline
$N$ & $b$     &  Slope$/\sqrt{\sigma}$ &  $\bar \chi$\\
\hline
169 & 0.355 &  19.54(25) & 0.76  \\
169 & 0.360 &  18.79(41) & 1.07   \\
\hline
289 & 0.355 &  19.38(32) & 2.10  \\
289 & 0.360 &  19.61(33) & 0.61  \\
289 & 0.365 &  19.32(33) & 0.94  \\
289 & 0.370 &  20.71(54) & 0.64  \\
\hline
361 & 0.355 &  19.07(45) & 2.87  \\
361 & 0.360 &  18.98(31) & 1.71  \\
361 & 0.365 &  18.82(21) & 1.10  \\
361 & 0.370 &  20.06(36) & 1.07  \\
\hline
    \end{tabular}
  \caption{Slope, in units of the string tension, of the linear dependence of the pion mass square on the twisted mass parameter $\mu$.  }
    \label{table4}
  \end{center}
\end{table}

\begin{table}[t]
  \begin{center}
    \begin{tabular}{|c|c|c|c|c|c|}
\hline
$N$ & $b$  &  $m_{\pi*}^{(0)}/\sqrt{\sigma}$ (W)& $\bar \chi$& $m_{\pi*}^{(0)}/\sqrt{\sigma}$ (Tw) & $\bar\chi$\\
\hline
289&0.355&3.42(26)&0.22&4.45(56)&0.12 \\
289&0.360&4.00(42)&0.16&4.16(51)&0.10 \\
289&0.365&4.70(18)&0.28&4.67(27)&0.10 \\
289&0.370&4.48(20)&0.28&4.45(38)&0.10 \\
\hline
361&0.355&3.40(46)&0.13&4.17(69)&0.18 \\
361&0.360&3.60(35)&0.10&4.62(30)&0.10 \\
361&0.365&4.32(38)&0.39&4.36(16)&0.13 \\
361&0.370&5.02(31)&0.10&5.09(33)&0.13 \\
\hline
    \end{tabular}
  \caption{Masses of  first excited state in the pseudoscalar channel extrapolated to the chiral limit for Wilson (W) and twisted mass (Tw) fermions.
  }
    \label{table_pion2}
  \end{center}
\end{table}

\begin{figure}[t]
\centering
\includegraphics[width=\linewidth]{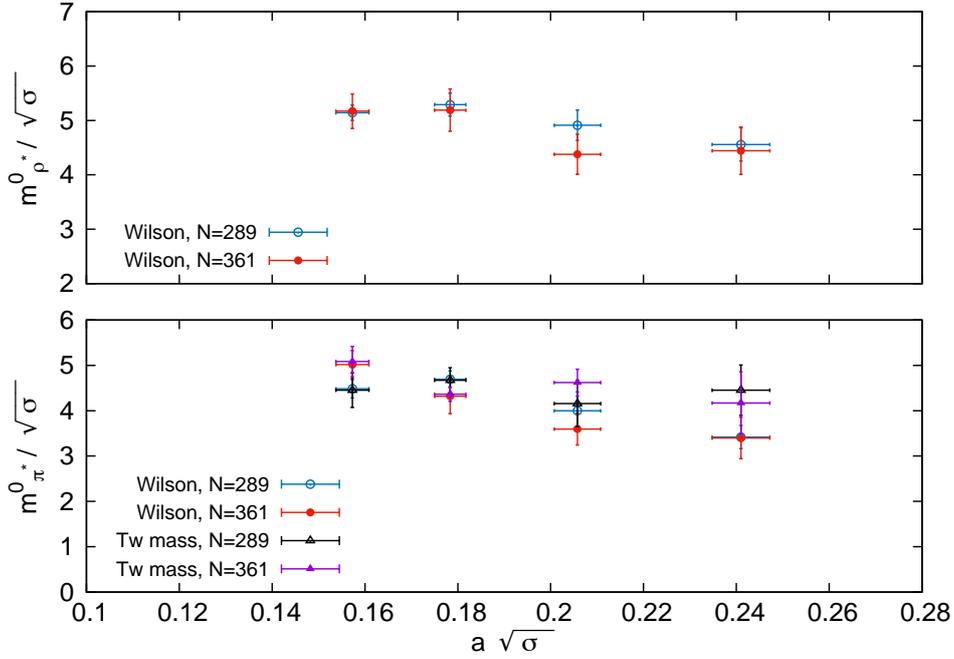}
\caption{Lattice spacing dependence of the first excited states in the pseudoscalar and vector channels, extrapolated to the chiral limit. 
}
\label{fig_m2_scaling}
\end{figure}

Having determined the mass of the pion we can also look at the masses
of the excited states with the same quantum numbers. The results of the GEVP for the first excited states turn out not to be precise enough and we have opted for estimating the mass of the excited states by  performing double exponential fits to the correlators, keeping the ground state mass fixed. The results for Wilson and twisted mass fermions are presented in tables~\ref{t:mpi_w} and ~\ref{t:mpi_tw}.
An extrapolation  to the chiral limit  using a linear dependence in the PCAC mass, at given $b$ and $N$, works well and gives the intercepts presented in table~\ref{table_pion2}, which have rather large statistical errors. For each value of $b$ we have 4 determinations depending on the methodology (Wilson or twisted mass) and the value of $N$, which are displayed as function of the lattice spacing in fig.~\ref{fig_m2_scaling}. This can give an idea of the systematic errors. Given that they are quite large, we cannot perform a reliable continuum extrapolation of the results, a constant fit to all the data gives for instance $m_{\pi^*}^{(0)}/\sqrt{\sigma}=4.4(2)$ with a large value of $\bchi=1.6$. Further work would be required to improve these results.

\subsubsection{Pion decay constant}
Another very interesting quantity to study is the pion decay constant
$f_\pi$. This is given in terms of  the  matrix element of the axial current between the vacuum and the pion states. In contrast with what happens with masses, this quantity is sensitive to normalization and renormalization of the operator involved. The naive extension of the standard QCD definition  diverges in the large $N$ limit as $\sqrt{N}$. Thus, as previous authors, we define the large $N$ decay constant as follows
\be
\label{fpidef}
f_\pi= \sqrt{2} F_\pi= \sqrt{\frac{3}{N m_\pi^2} }\langle 0 | A_0(0) | \pi; \vec{p}=0 \rangle 
\ee
where $A_0$ is the temporal component of the axial vector current. The definition is cooked to coincide with the standard one for $N=3$. As it stands, the definition is symbolic  if we do not specify how the  $| \pi;\vec{p}=0 \rangle$ state is normalized. The formula assumes the relativistically invariant normalization $\langle p | q \rangle = (2 \pi)^3 (2 p_0) \delta^3(\vec{p}-\vec{q})$.

In practice the matrix element appearing in Eq.~\eqref{fpidef}  can be extracted from the correlation function of the temporal component of the axial current with any operator that acts as a pion interpolating field. Dividing out by the square root of the correlation function of the interpolating field with itself, we eliminate the dependence on the choice of interpolating field. The mechanism is true both on the continuum as on the lattice. As a lattice interpolating field we can use the optimal operator selected by the GEVP method, reducing the possible contamination with excited states. In any case, this can be monitored by verifying that the correlator decreases in time as $\exp(-t m_\pi)$. This is essentially the same method used by other authors as for example Ref.~\cite{\BALIMESONS}. A final word of warning comes from the aforementioned necessity of normalizing and renormalizing the lattice Axial current operator to match with the continuum definition. The normalization is rather standard. With our choice of the Wilson Dirac operator the lattice quark field is given by $\Psi_L= \sqrt{a^3/(2 \kappa)} \Psi$. In addition the Axial current is a composite operator and gets a renormalization factor $Z_A$. This depends on the choice of lattice discretized operator. Here we restrict ourselves to the ultralocal version 
$A_0^L(x)= \bar{\Psi}_L(x)\gamma_0 \gamma_5 \Psi_L(x)$. Finally, we can express the lattice spacing and the pion mass in string tension units to determine $F_\pi/(\sqrt{\sigma} Z_A)$ for each of our datasets.  The results are collected in the Table~\ref{t:fpi_wilson}.

Do our data scale? Examining  the results for N=289, one can see that if we compare data points having the same value of the pion mass, the tendency is that  $F_\pi/(\sqrt{\sigma} Z_A)$ tends to decrease with increasing $b$. For example the data at $b=0.36$, $\kappa=0.1577$ has very similar pion mass to that of $b=0.365$ and $\kappa=0.1562$, but its value of $F_\pi/(\sqrt{\sigma} Z_A)$ is 10\% higher.  However, one cannot deduce from this a sizable violation of scaling, since the renormalization factor $Z_A$ introduces a explicit $b$-dependence of the result which goes in the same direction. Hence, to verify this one would need to know the value of $Z_A$ for each $b$. 

Driven by this problem we decided to try a different method for determining the value of $F_\pi$\footnote{We thank our colleagues Carlos Pena and Fernando Romero-L\'opez for discussions about this point}. This is actually the main reason why we decided to include twisted mass fermions. One of the advantages of this method is that it allows a determination of $F_\pi$ without a renormalization factor~\cite{DellaMorte:2001tu}. Furthermore, the procedure is somewhat different and serves as an additional test of the robustness of our results. The value of $f_\pi$ is extracted from the expression 
\be
 f_\pi=  \frac{ 2 \mu \sqrt{3}}{m_\pi^2\sqrt{N} } \langle 0 | \bar\Psi(x) \gamma_5 \Psi(x) | \pi;  \vec{p}=0\rangle
\ee
Deriving the appropriate reduced lattice formula poses no problem. Again, the matrix element is extracted from the correlation function of the ultralocal pseudoscalar operator with the optimal one. The propagator is now the inverse of the twisted mass one, given in the previous section. This includes the Wilson-Dirac operator set at the value of  $\kappa_c$ determined earlier for Wilson fermions. Namely the value at which the extrapolated PCAC mass vanishes. The results then depend on the additional parameter $\mu$, for which we choose 4 values  for each data set. As mentioned earlier, the pion mass square obtained in each case vanishes  linearly in $\mu$. The corresponding  results  for $F_\pi$ are given in Table~\ref{t:fpi_tw}. 

\begin{figure}[t]
\centering
\includegraphics[width=\linewidth]{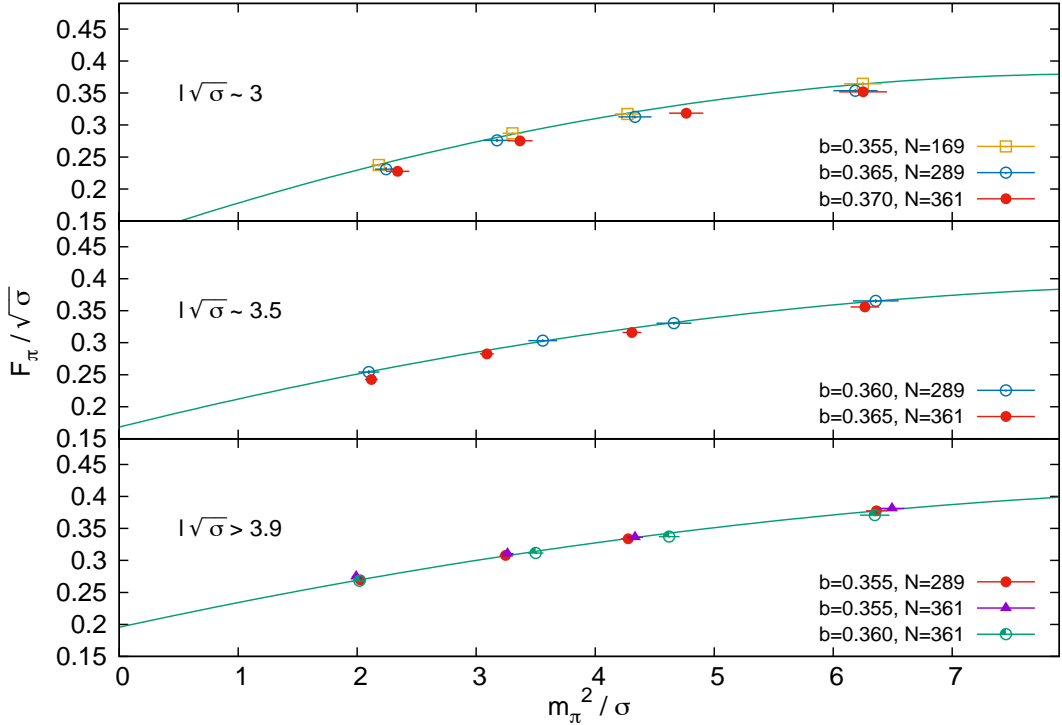}
\caption{Dependence of the pion decay constant on the pion mass square at approximately fixed physical volume for twisted mass fermions.
}
\label{figfpitm}
\end{figure}

We still observe that some of the results having similar values of $m_\pi$ have non-compatible  values of $F_\pi$. However, after analyzing the results in detail, we discovered that this phenomenon is due to a sizable dependence on the effective volume. In Fig.~\ref{figfpitm} we display the values of $F_\pi$ as a function of the pion mass square arranged into sets having similar values of the effective  box size $a\sqrt{N}= l \sqrt{\sigma}$. The data scale well within errors. However, the values of $F_\pi$ tend to decrease sizably for small volumes. Having results for three different values of $N$ has been crucial in discovering the origin of this disagreement.

\begin{figure}[t]
\centering
\includegraphics[width=\linewidth]{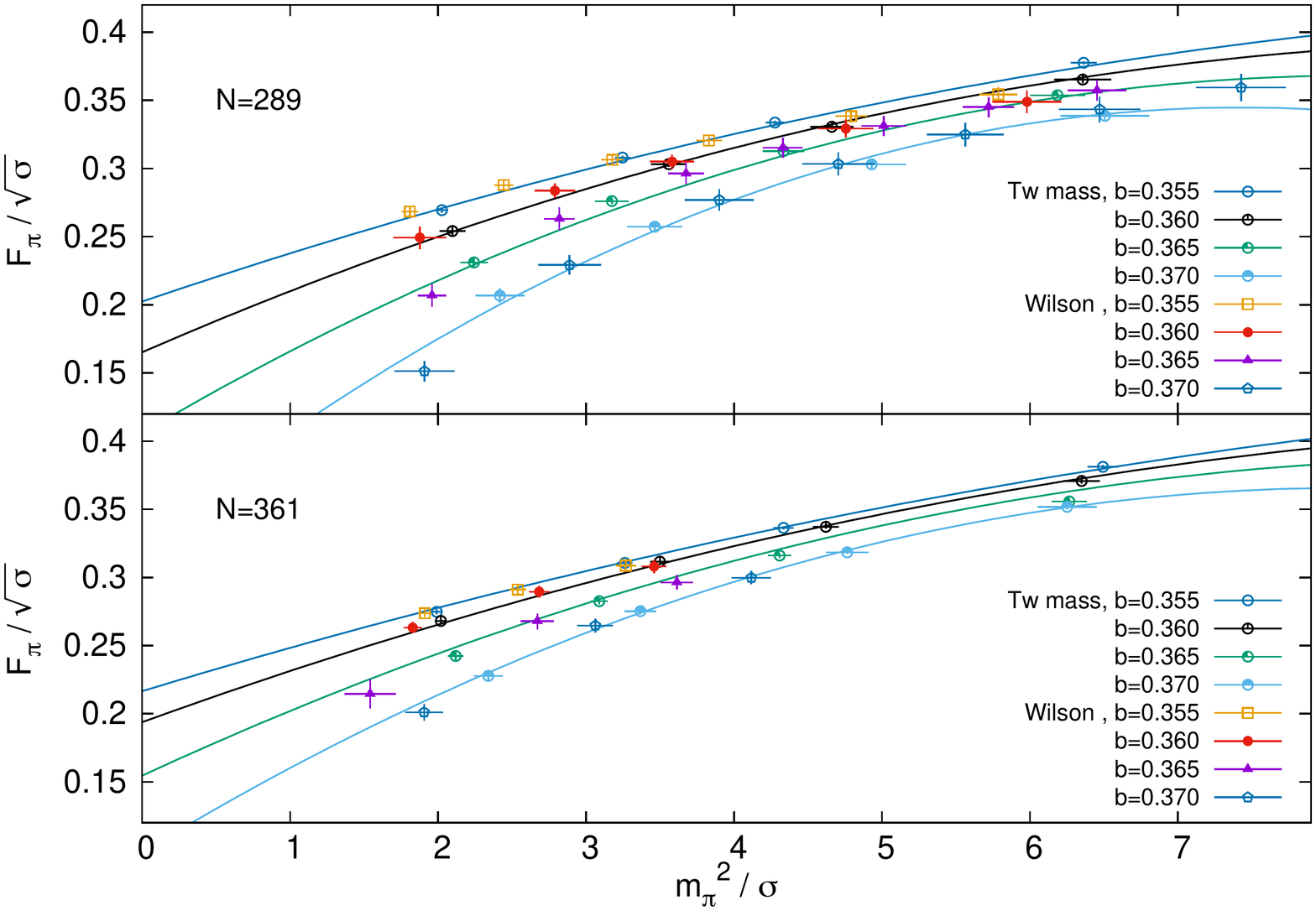}
\caption{Dependence of the pion decay constant on the pion mass square.
The lines correspond to the function $F(m_\pi) + B(m_\pi)  \exp\{-C \sqrt{\sigma} l\}$, with $F(m_\pi) = 0.232(6) + 0.029(3) m^2_\pi /\sigma -0.0008(3) m^4_\pi /\sigma^2 $, 
$B(m_\pi) = -7.9(2.5) + 2.2(7) m^2_\pi /\sigma -0.19(7) m^4_\pi /\sigma^2 $, and $C=1.36(11)$. The coefficients have been extracted from a simultaneous fit of all the twisted mass data with $N=289$ and $N=361$, the value of $\bar \chi$ for the fit is 0.85. }
\label{f:fpi_ldep_tot}
\end{figure}

In order to get maximal information from our data we have tried to parameterize the volume dependence of the results by fitting the data to an exponentially decreasing  function of the effective size: $F_\pi(m_\pi) + B(m_\pi)\, \exp\{-C \sqrt{\sigma} l\}$. The coefficient $C$ turns out to be 1.36(11) which explains why our results obtained for relatively small volumes are still sizably dependent on the volume. Curiously we did not find any significant  dependence in the determination of the masses. The fitted function 
$F_\pi(m_\pi)$ gives our estimate of the pion decay constant at infinite volume and infinite $N$. In the region covered by our data it can be described well by a second degree polynomial in $m_\pi^2$. The same goes for the coefficient $B(m_\pi)$ of the exponentially decreasing term. Fig.~\ref{f:fpi_ldep_tot} shows all our data for $N=289$ and $N=361$ together with the solid lines giving the result of the fit. An important quantity is the value of $F_\pi$ in the chiral limit. We have done various fits with different parameterizations and in all cases it seems that a value of $F_\pi(0)=0.22(1)(2)$ is a safe estimate. The first error being statistical and the second systematic.

Once we have obtained $F_\pi$ from the twisted mass data we can come back to analyze the results obtained for Wilson fermions. Knowing the value of $F_\pi$ the Wilson data allows a non-perturbative determination of $Z_A$. Unfortunately, this cannot be done point by point since the twisted mass and Wilson data do not correspond to exactly the same values of $m_\pi^2$. However, it is highly non-trivial that all the Wilson fermion data can be well described by the same functional dependence on $m_\pi^2$ up to a multiplicative factor $1/Z_A$. Remarkably it can, as shown in Fig.~\ref{f:fpi_ldep_tot}, where the twisted mass data are displayed together with the Wilson Dirac data after a suitable rescaling.

As a spin-off we have obtained a non-perturbative determination of $Z_A$. The corresponding values are listed in Table~\ref{table_ZA}.  This can be compared with the value  computed in perturbation theory to two loop order in Ref.~\cite{Skouroupathis:2008mf} giving
\be
\label{ZA_PERT}
Z_A= 1- 0.03735 \lambda -0.00058 \lambda^2 
\ee
It is well-known that truncated perturbative results on the lattice give poor estimates when computed in terms of the bare coupling constant $\lambda=1/b$. Much better results follow when using improved couplings $\lambda_I$ of which there are various examples. In Table~\ref{table_ZA} we use various customary definitions to  re-express the perturbative formula Eq.~\ref{ZA_PERT} in terms of those improved couplings. In general, the perturbative estimates tend to give higher values than our non-perturbative result. Remarkably the same phenomenon takes place for SU(3) as commented in Ref.~\cite{\BALIMESONS}. 

\begin{table}[t]
  \begin{center}
    \begin{tabular}{|c||c|c|c|c|c|}
\hline
& {\small Non-perturb.} & {\small Non-perturb.} & {\small Perturb. } $\lambda_I=$& {\small Perturb.} $\lambda_I=$ &{\small Perturb.} $\lambda_I=$  \\
b & N=289 & N=361 & $1/(bP(b))$ & $-8 \log(P(b))$  & $8 (1-P(b))$\\ \hline
\hline
0.355 &   0.807(7) &  0.806(4) & 0.9161 &   0.8960 &   0.8766  \\ \hline
0.360 &  0.827(10) &  0.812(6)& 0.9154 &   0.8971 &   0.8797 \\ \hline
0.365  & 0.854(8)  & 0.801(10)&  0.9149 &   0.8983 &   0.8825\\ \hline
0.370  & 0.85(2)   & 0.81(1) & 0.9147 &   0.8994 &   0.8849 \\ \hline
    \end{tabular}
  \caption{The first and second columns contain our non-perturbative determination of $Z_A$ from the $N=289$ and $N=361$ data. The remaining columns contain two-loop perturbative predictions using  various definitions of the improved coupling $\lambda_I$ in terms of  the plaquette value $P(b)$.}
    \label{table_ZA}
  \end{center}
\end{table}

%% file: vector_channel.tex
\begin{figure}[t]
\centering
\includegraphics[width=0.75\linewidth]{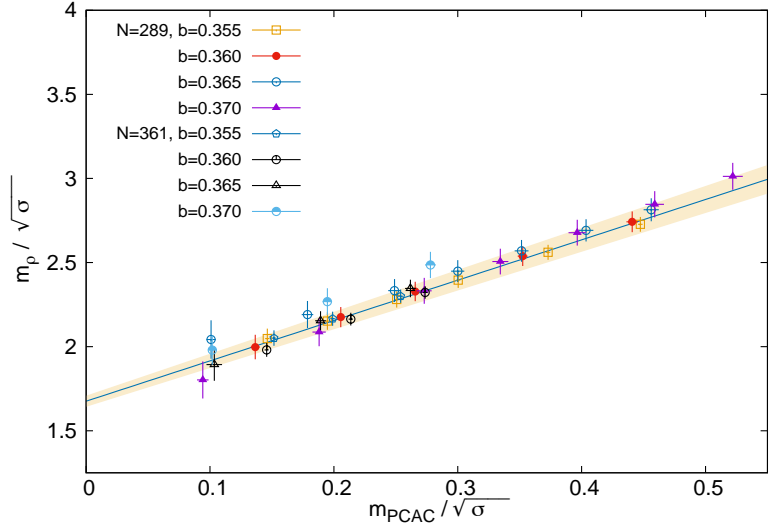}
\caption{Dependence of the vector meson mass as a function of the PCAC mass, all in units of the string tension. 
}
\label{fig_rho}
\end{figure}

\begin{figure}[t]
\centering
\includegraphics[width=0.75\linewidth]{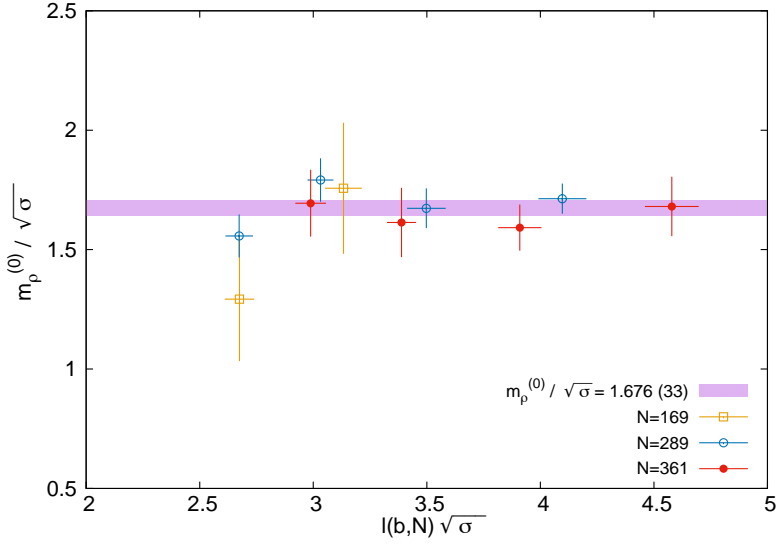}
\caption{Finite size dependence of the chirally extrapolated vector meson mass, $m_\rho^{(0)}/\sqrt{\sigma}$.  The purple band in the plot is the chiral extrapolation displayed in fig.~\ref{fig_rho}, determined by a joint fit of the PCAC mass dependence of the data with $l(b,n) \geq 3.4 $.
}
\label{fig_rho_voldep}
\end{figure}

\begin{table}[ht]
  \begin{center}
    \begin{tabular}{|c|c|c|c|c|}
\hline
$N$ & $b$  &  Slope & $m_\rho^{(0)}/\sqrt{\sigma}$& $\bchi$\\
\hline
169 & 0.355 & 1.2(1.6) & 1.76(27) & 0.44  \\
169 & 0.360 & 3.2(1.5) & 1.29(26) & 0.08   \\
\hline
289 & 0.355 & 2.27(20)  & 1.713(63) & 0.05 \\
289 & 0.360 & 2.44(27)  & 1.673(84) & 0.12\\
289 & 0.365 & 2.22(27)  & 1.792(91) & 0.16 \\
289 & 0.370 & 2.81(25)  & 1.557(90) & 0.13 \\
\hline
361 & 0.355 & 2.44(59)  & 1.68(12)  & 0.001 \\
361 & 0.360 & 2.67(44)  & 1.59(10) & 0.03 \\
361 & 0.365 & 2.81(67)  & 1.61(15)  & 0.20 \\
361 & 0.370 & 2.87(67)  & 1.69(14)  & 0.21 \\
\hline
\hline
$l(b,N) \geq 3.4 $ & &  2.40(12) &1.676(33) & 0.44 \\
\hline
    \end{tabular}
  \caption{Slope and intercept of the chiral extrapolation of the vector meson mass. The last row is determined by a joint fit of all the data with  $l(b,n) \geq 3.4 $.  }
    \label{table_rho}
  \end{center}
\end{table}

\begin{table}[ht]
  \begin{center}
    \begin{tabular}{|c|c|c|c|c|c|}
\hline
$N$ & $b$  &  $m_{\rho*}^{(0)}/\sqrt{\sigma}$ (W)& $\bchi$\\
\hline
289&0.355&4.55(31)&0.22 \\
289&0.360&4.91(28)&0.16 \\
289&0.365&5.29(21)&0.28 \\
289&0.370&5.14(14)&0.28 \\
\hline
361&0.355&4.44(43)&0.1\\
361&0.360&4.38(37)&0.1 \\
361&0.365&5.19(39)&0.18 \\
361&0.370&5.17(32)&0.15 \\
\hline
    \end{tabular}
  \caption{Intercept of the chiral extrapolation of the first excited state in the vector channel.}
    \label{table_mrho2}
  \end{center}
\end{table}

 We have also analyzed the lowest lying spectrum in the vector channel using in this case Wilson fermions. The analysis, following the strategy presented in sec.~\ref{s:mass_comp}, has allowed a determination of the  $\rho$ mass, as well as the mass of the first excited state in this channel, $\rho^*$.  The final results for all our lattices are compiled in table~\ref{t:mrho} in Appendix~\ref{tables_wf}.

The value of the $\rho$ mass shows a linear dependence in the PCAC quark mass, as expected at leading order in chiral perturbation theory in the large $N$ limit~\cite{Booth:1996hk}. This dependence is displayed in fig.~\ref{fig_rho} for all our lattices. By performing a simultaneous fit to all the points with physical size $l(b,n) \geq 3.4 $, we obtain an intercept in the chiral limit given by $m_\rho^{(0)}/\sqrt{\sigma}=1.676(33)$,  fitting instead all points leads to a compatible value $m_\rho^{(0)}/\sqrt{\sigma}=1.634(26)$, slightly worsening the value of $\bchi$ from 0.44 to 1.0. We have also performed independent fits to each data set, keeping $b$ and $N$ fixed. The corresponding intercepts and slopes are given in table
~\ref{table_rho}. In fig.~\ref{fig_rho_voldep}, the values of the intercepts are displayed as a function of the physical volume, $l(b,N)$; with the exception of the smallest lattice volume, there is no significant finite volume effect in the results. The scaling with the lattice spacing is very good and we see no clear lattice spacing dependence within the estimated errors. 

\begin{figure}[t]
\centering
\includegraphics[width=0.75\linewidth]{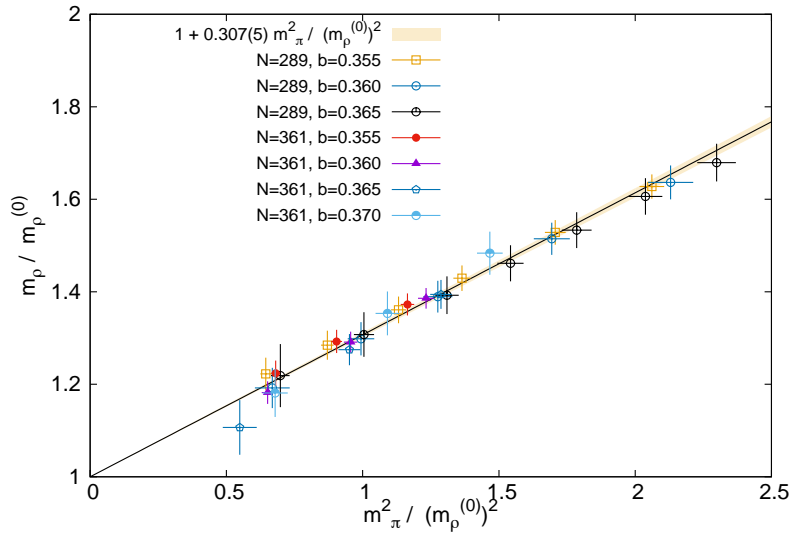}
\caption{Dependence of the $\rho$ mass on the pion mass square, both normalized in units of the value of the $\rho$ mass in the chiral limit $m^{(0)}_\rho$. 
}
\label{fig_edimb_rho}
\end{figure}

A common way of displaying the dependence of the $\rho$ mass on the quark mass is to use the value of $m_\rho^{(0)}$ to set the scale and monitor the dependence of $m_\rho/m_\rho^{(0)}$ on $(m_\pi/m_\rho^{(0)})^2$, this eliminates renormalization ambiguities in the definition of the PCAC quark mass and may get rid of some of the systematic errors in the determination of the scale. All our results for $N=289$ and $N=361$, with $l(b,N) \sqrt{\sigma} > 2.9$, are plotted in this way on fig.~\ref{fig_edimb_rho}. The dependence on the pion mass square is linear with a slope given by 0.307(5) and with $\bchi = 0.45$ .

We have as well determined the masses of the first excited state in the vector channel, following the same procedure discussed for the pseudoscalar. The results are presented in table~\ref{t:mrho}. The intercepts of the linear extrapolation to the chiral limit are given
in table~\ref{t:mrho} and displayed as a function of the lattice spacing in fig.~\ref{fig_m2_scaling}. Assuming that the lattice spacing dependence is within errors, and fitting all the results to a constant gives a mass for the excited state of 5.0(2) in units of the string tension with a value of $\bchi=1.2$.

%% file: heavy_states.tex
\begin{figure}[t]
\centering
\includegraphics[width=0.73\linewidth]{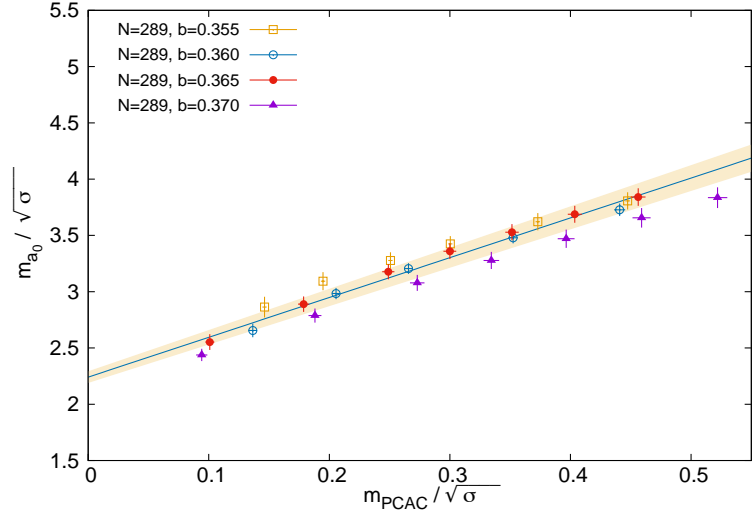}
\caption{Dependence of the $a_0$ meson mass as a function of the PCAC mass, all in units of the string tension. The yellow band corresponds to a joint fit of the $b=0.360$ and $b=0.365$ data sets.
}
\label{fig_a0}
\end{figure}
\begin{figure}[t]
\centering
\includegraphics[width=0.73\linewidth]{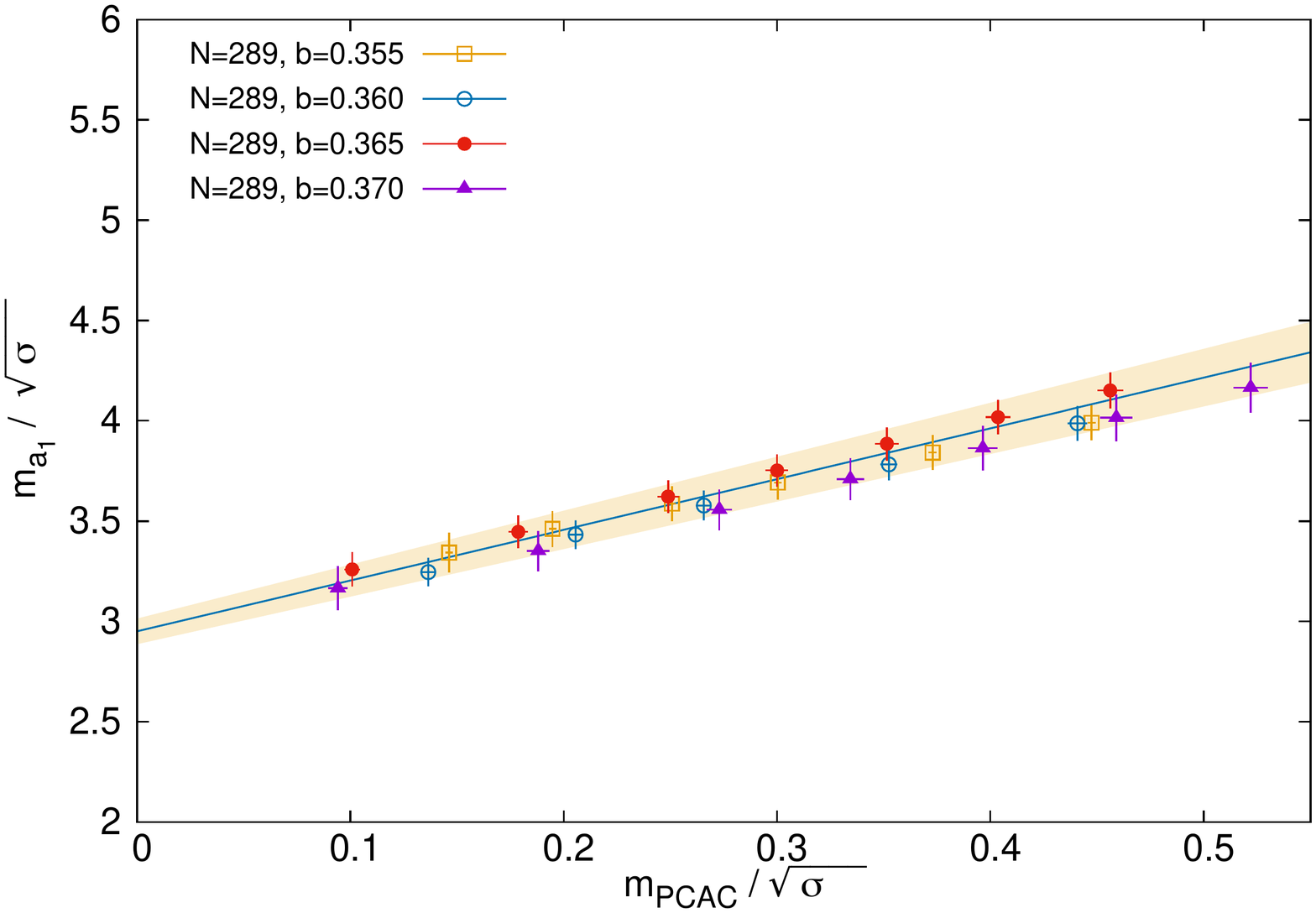}
\caption{Dependence of the $a_1$ meson mass as a function of the PCAC mass, all in units of the string tension. The yellow band corresponds to a joint fit of the $b=0.360$ and $b=0.365$ data sets.
}
\label{fig_a1}
\end{figure}

\begin{figure}[t]
\centering
\includegraphics[width=0.73\linewidth]{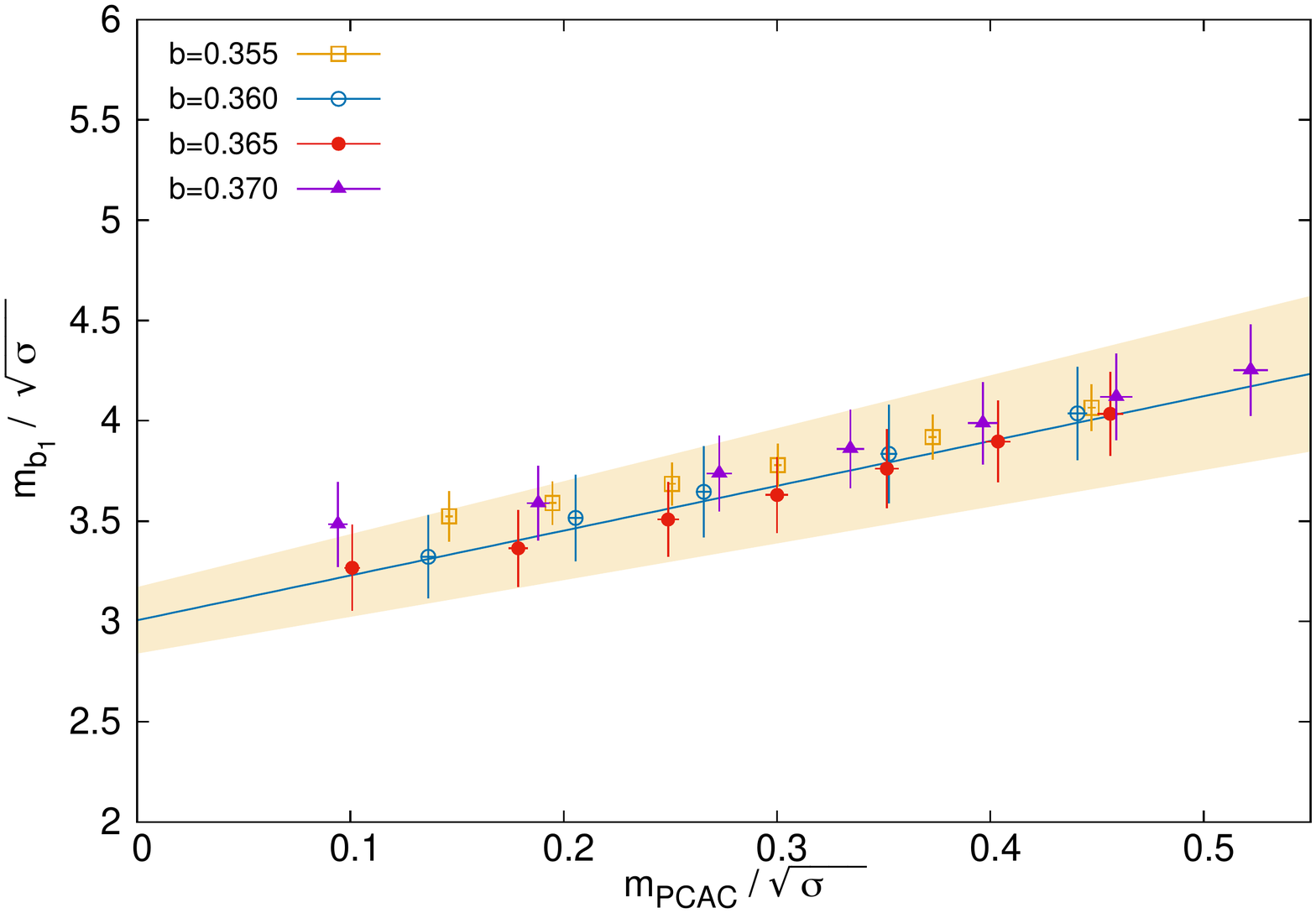}
\caption{Dependence of the $b_1$ meson mass as a function of the PCAC mass, all in units of the string tension. The yellow band corresponds to a joint fit of the $b=0.360$ and $b=0.365$ data sets.
}
\label{fig_b1}
\end{figure}

 The ground-state masses in the  $a_0$, $a_1$ and $b_1$ meson channels, for $N=289$ and various lattice spacings, are presented in table~\ref{t:mheavy} in Appendix~\ref{tables_wf}. They have been obtained using Wilson fermions. As mentioned in section
~\ref{s:mass_comp}, we extract the masses from a fit to the {\it optimal} correlator, including in this case operators with smearing up to 50 in the basis. This allows to determine the ground state masses with good accuracy, although our results are not precise enough to extract the masses of the excited states in each channel.
 
\begin{figure}[t]
\centering
\includegraphics[width=\linewidth]{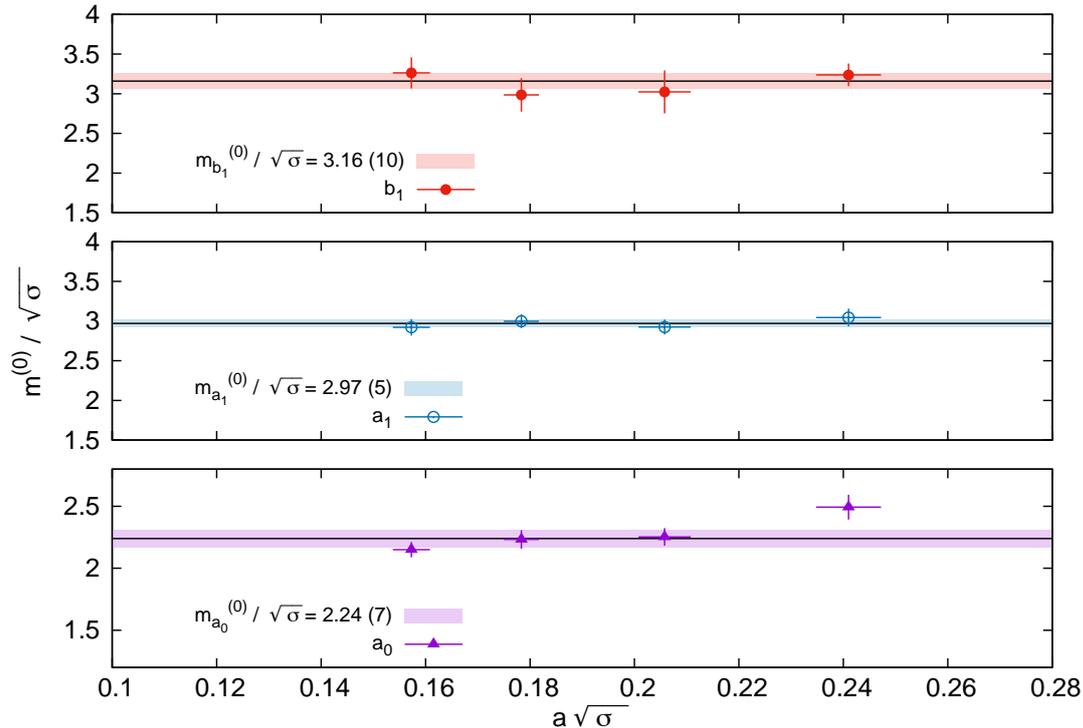}
\caption{Lattice spacing dependence of the chirally extrapolated $a_0$, $a_1$, and $b_1$ meson masses for $N=289$. The bands correspond to a simultaneous fit of all the results to a unique value, independent of the lattice spacing.}
\label{fig_mheavy_scaling}
\end{figure}

 As in the case of the $\rho$ meson,  the extracted masses depend linearly on the PCAC mass. Figures~\ref{fig_a0} - \ref{fig_b1} exhibit this linear dependence for the three different channels. A fit in each channel done jointly to the sets with $b=0.360$ and $0.365$ gives as intercepts in the chiral limit: $m^{(0)}_{a0}= 2.24(5) \sqrt{\sigma}$, $m^{(0)}_{a1}= 2.95(6) \sqrt{\sigma}$, and $m^{(0)}_{b1}= 3.01(17) \sqrt{\sigma}$. The corresponding slopes are given in  table~\ref{table_heavy}, where we also collect the parameters resulting from fits done at a fixed value of the inverse lattice coupling $b$.  As shown in fig.
~\ref{fig_mheavy_scaling}, the results show good scaling towards the continuum limit.  The dependence on the lattice spacing is negligible in the case of the  $a_1$ and $b_1$ states, while the scalar meson $a_0$ shows a tendency to decrease towards the continuum limit. The bands in the plot are obtained from a constant fit to all the results, which works well in all cases except for the $a_0$ mass. This issue will be further discussed in section~\ref{s:discussion}, where we will present our final estimates for the continuum extrapolated masses.

\begin{table}[t]
  \begin{center}
    \begin{tabular}{|c|c|c|c|c|c|c|c|c|}
\hline
$N$ & $b$  &  $m_{a_0}^{(0)}/\sqrt{\sigma}$& Slope &  $m_{a_1}^{(0)}/\sqrt{\sigma}$& Slope &  $m_{b_1}^{(0)}/\sqrt{\sigma}$&Slope & $\bar \chi$\\
\hline
289 & 0.355 & 2.49(10)&3.01(32) &3.04(11) & 2.13(36) &   3.23(14)& 1.83(47) & 0.5/0.1/0.1  \\
289 & 0.360 & 2.25(7) &3.44(24) &2.92(9)  &2.43(33) & 3.02(27)&2.31(93) &0.9/0.1/0.1\\
289 & 0.365 & 2.23(7) &3.64(24) &3.00(9)  &2.52(28) & 2.99(21) &2.23(68) &0.5/0.1/0.2\\
289 & 0.370 & 2.15(6) &3.31(20) &2.92(10) &2.37(31) & 3.26(20) &1.84(58) &0.4/0.1/0.2\\
\hline
289 &  &  2.24(5) &3.54(17) &  2.95(6) & 2.53(21) & 3.01(17) & 2.23(55) & 0.7/0.6/0.2 \\
\hline
    \end{tabular}
  \caption{Slope and intercept of the chiral extrapolation of the  masses in the $a_0$, $a_1$, and $b_1$ meson channels. The last row is determined by a joint fit of the data with $b=0.360$ and 0.365.  }
    \label{table_heavy}
  \end{center}
\end{table}

%% file: discussion.tex
In this section we will summarize our results and analyze them  in relation with other estimates and expectations of meson masses at large $N$. We have provided a lot of information in tables which could help other researchers to  draw their own conclusions. Our methodology has the advantage of having negligible standard large $N$ corrections. However, other type of large $N$ corrections arise, taking  the form of finite volume corrections. This can be understood and avoided by working at large enough effective volumes in physical units. Indeed, after monitoring for this effect we do not see appreciable dependence in the masses. On the contrary, the effect is quite sizable in the pion decay constant $f_\pi$, and extrapolation is needed to obtain sensible results. Fortunately, in this case we have two independent methods (Wilson and twisted mass fermions) to provide additional support to our analysis. 

To obtain the physical masses from lattice computations one has to extrapolate to the continuum limit.  We have expressed the masses obtained at all lattice spacing values in units of the string tension, so that in the continuum limit they should approach a constant value, which is the final continuum result. All the meson masses, with the exception of  the pion mass, depend linearly on the quark mass. On the other hand spontaneous chiral symmetry breaking implies  that the pion mass square also depends linearly on the quark mass. For a more physical comparison 
we preferred to combine the two previous results and express the meson masses at each lattice spacing as follows
\be
\label{finalfit}
\frac{m_X}{\sqrt{\sigma}}= \frac{m_X^{(0)}}{\sqrt{\sigma}} + Y_X \frac{m_\pi^2}{\sigma}
\ee
where $X$ specifies the meson in question. Indeed, this is precisely what we see. However,  the value of the masses in the chiral limit $\frac{m_X^{(0)}}{\sqrt{\sigma}} $ at each lattice spacing value,  were obtained  by extrapolating to vanishing PCAC mass, since that quantity is more precise and less affected by finite effective volume corrections than the pion mass itself. Once that is fixed, one can determine the  slope coefficient $Y_X$ from the remaining data. For the case of $F_\pi$ our results can be well fitted with the  following simple parameterization:
\be
\frac{F_\pi^2}{\sigma} = \frac{(F_\pi^{(0)})^2}{\sigma} +  Y_F \frac{m_\pi^2}{\sigma}
\ee
In this case a large $N$ extrapolation is also needed as explained in the previous section. 

The final stage is the extrapolation of the chiral limit meson masses and slopes $Y_X$ to the continuum limit. Our lattice results, obtained for four different values of the lattice spacing, have to be extrapolated to vanishing spacing.  In general,   to the leading  approximation, the   continuum value is  approached linearly or quadratically in the lattice spacing depending on the quantity, the system under study and particular discretization employed. For Wilson fermions we expect a linear approach. However, the   majority of our physical observables are consistent with a negligible small slope compared to the statistical errors.  This can be seen from Figs.~\ref{fig_rho_voldep},~\ref{fig_mheavy_scaling}.
Indeed, fitting the values obtained for masses and slopes for all lattice spacings to a unique value gives values of  $\bar{\chi}$ which are well below 1, signalling a good fit. In accordance with that, we have given as  our best estimates of the continuum values the ones  obtained by this  simultaneous fit. These values are collected in Table~\ref{t:finaltable}. The described  situation applies for  all the slopes $Y_X$ as well as the masses of the   $\rho$, $a_1$ and $b_1$ mesons  in the chiral limit.  The only exception is the value of the $a_0$ mass. In this case the fit is not statistically favourable giving a value of $\bar{\chi}=1.7$. More importantly, the data shows a systematic trend towards a decrease with the lattice spacing. Thus, our best estimate for the mass of $a_0$ will follow from the procedure explained below. 

 In any case, it is not possible to exclude a possible linear or quadratic dependence on the lattice spacing even for the cases in which the data is consistent with a constant value. The extrapolation of the data using linear or quadratic dependencies on the lattice spacing also depends on which observable ($\sqrt{\sigma}$, $\bar{r}$ or $t_0$) is used to fix the scale, despite the good agreement seen in table~\ref{tableconf}. For the case of the $\rho$, $a_1$ and $b_1$ meson masses, the different extrapolations do not ameliorate the $\bar{\chi}$ of the constant fit. However, we made use of the span defined by all the  different  extrapolations to give an estimate of  the systematic errors, since they exceed those  coming from other sources. This shows  the typical problem involved in extrapolations. The boundaries of the range covered by the different continuum limit extrapolations relative to the best statistical average define a positive and a negative  value, shifting up or down the best value. These shifts are shown in Table~\ref{t:finaltable} as a column vector of values to be added or subtracted to the last significant digits of the best value.
 
 The situation for the  $a_0$ mass is quite different. In this case, all the extrapolations based on the different choices  of scale-setting predict a decrease with the lattice spacing $a$, accompanied with a better statistical significance, signalled by values of $\bar{\chi}$ of order 0.75-0.9.  The continuum limit values obtained by fitting a linear dependence in $a/\bar{r}$ and $a/\sqrt{8 t_0}$ are quite consistent with each other. After converting them back to string tension units using $\bar{r}\sqrt{\sigma}=1.035$ and $\sqrt{8 t_0\sigma}=1.078$ we get $m_{a_0}^{(0)}/\sqrt{\sigma}=1.829$ and $1.816$ respectively. Hence, we take the weighted average as our best estimate shown on Table~\ref{t:finaltable}. The systematic errors are depicted in the same fashion ranging from the constant fit to the linear fit using $a\sqrt{\sigma}$.
 
 In the case of the  slopes $Y_X$ (shown  in Table~\ref{t:finaltable}) the systematic errors associated to the continuum limit extrapolations lie well within  the range of statistical errors. Our best estimates for the  pion decay constant and its corresponding slope are given,  followed by the statistical and systematic error. The latter also depends on the large $N$ extrapolation presented in the previous section.  

\begin{table}[t] 
\begin{tabular}{||c||c|c|c|c|c||}
\hline \hline
$X$  & $\rho$  &  $a_1$ & $b_1$ & $a_0$ & $F_\pi$ \\ \hline \hline
$\frac{m_X^{(0)}}{\sqrt{\sigma}}$ & $1.67(3)\begin{pmatrix}+9 \cr -18\end{pmatrix}$ & $2.97(5)\begin{pmatrix}+8 \cr -18\end{pmatrix}$ & $3.16(10)\begin{pmatrix}+11 \cr -17\end{pmatrix}$& $ 1.83(15) \begin{pmatrix}+41 \cr -24\end{pmatrix}$ & $0.215(5)(20)$\\ \hline
$Y_X$ & 0.19(1)& 0.18(1) &0.15(2) & 0.25(2)&0.0159(2)(20)\\ \hline \hline
\end{tabular}
\caption{Masses of mesons in the chiral and continuum limit and slopes corresponding to Eq.~\eqref{finalfit}. The central value is our best  statistically significant estimate, corresponding to a constant fit for the $\rho$, $a_1$ and $b_1$ mesons and a linear fit in $a(b)$ for the $a_0$ mass (see text). The column vector give the maximum positive and negative shifts of our best estimate needed to cover the range of all  different continuum limit extrapolations.  They can be interpreted as the  systematic errors of our continuum limit results (see text).}
\label{t:finaltable}
\end{table}

Now let us compare our results with other predictions and calculations. The first comparison can be made with QCD and its meson spectrum. Setting the string tension to 
$\sqrt{\sigma}=440 MeV$, the values of the physical I=1 meson masses in string tension units are given by 
1.76, 2.82, 2.82 and   2.23  for $\rho$, $a_1$, $b_1$ and $a_0$ respectively, which are not too far from the 1.71, 2.99, 3.175 and 1.855 of our large $N$ best estimates for the physical value of  $m_\pi^2/\sigma\sim 0.1$. Hence, it seems that, in what respects to meson masses,  large $N$ provides  a fairly  good approximation to the real world.

Our next comparison is  with other lattice determinations of the spectrum.  As mentioned in the introduction, results using the quenched approximation extrapolated to large $N$ started some years ago~\cite{DelDebbio:2007wk,Bali:2008an,DeGrand:2012hd}. Results based on ideas of volume independence similar to ours appeared even  earlier~\cite{Narayanan:2005gh,Hietanen:2009tu}. Very interestingly, some results have also been obtained with two flavours of dynamical quarks, with values reasonably close~\cite{DeGrand:2016pur}.  The most complete published lattice work is that of  Ref.~\cite{\BALIMESONS}. It involves  a very detailed and thorough work producing results on large $N$ spectroscopy by extrapolation of the results obtained at various $N$, including some   at $N=17$. These results are obtained at a unique value of the lattice spacing, so that an analysis of the lattice spacing dependence  is not possible. However, since the authors use Wilson fermions and a Wilson action with a  value of the coupling which seems to correspond to b=0.36 at large $N$, we can compare their results with what we obtain at that same coupling. The chiral limit masses for $a_1$ and $b_1$ and $F_\pi$ are perfectly consistent within errors. Our result for the $\rho$ in the chiral limit (1.63(8))  is slightly larger than their result (1.54(1)), while for the $a_0$ it is the other way round (2.25(7) versus 2.40(3)). In any case, there is a  very good qualitative agreement, which is remarkable  given the very different methodology used by the two determinations. Incidentally our results on the rho mass and $F_\pi$ do not agree with those of Refs.~\cite{Narayanan:2005gh,Hietanen:2009tu}, despite they being closer methodologically to ours.

An effort to obtain results in the continuum limit  was done by some members of the same collaboration. In particular a detailed study with 4 lattice spacing (like ours) for SU(7)  has appeared in proceedings of  conferences~\cite{Bali:2013fya}. A more detailed chart including an estimate of the continuum extrapolation of meson masses at large $N$ was presented in the thesis dissertation of Luca Castagnini~\cite{Castagnini:2015ejr}\footnote{We thank Marco Bocchicchio and Gunnar Bali for pointing us to this reference}.
The study covers 4 values of the lattice spacing with the finer lattices similar to ours and one point coarser than our data. 
The final results involve a double extrapolation to $N=\infty$ and $a=0$, which is achieved by a 4 parameter fit to the data of each observable. Hence, the results have to be taken cum granum salis, as recognized by the author. Nonetheless, the final table is strikingly similar to our best values giving chiral limit masses of 1.687(24), 2.93(11), 2.97(13) for $\rho$, $a_1$ and $b_1$ respectively and $F_\pi=0.197(20)$. The slopes in $a$ obtained by the fit for $a_1$ and $b_1$ are small, in numerical agreement with our results which are consistent with vanishing slopes within errors. There is certain disagreement in their predicted slope in $a$ for the $\rho$. Mysteriously we end up  having the same estimate  for the rho  mass in the chiral limit and large $N$.   

The case for the scalar meson  $a_0$ deserves special attention. Both our results as well as those of Refs.~\cite{Bali:2013fya,Castagnini:2015ejr} point towards  a sizable positive
slope with $a\sqrt{\sigma}$. Our best estimate for the $a_0$ mass in the chiral limit 1.83(15)
is remarkably close to the value  1.81(17) given in Ref.~\cite{Castagnini:2015ejr}. This
agreement  gives a stronger evidence that our observed a-dependence is not a statistical fluctuation but rather a genuine effect. It seems that the scalar mesons are rather special, also having a larger quark
mass dependence $Y_{a_0}$ than the remaining mesons. Indeed, the same happens in QCD and a long lived discussion has been centered about them (see
Ref.~\cite{Pelaez:2015qba} for a recent review). Actually, it has been
proposed that the study of the behaviour of the masses at large $N$
could help in settling some points~\cite{Pelaez:2006nj}. A
good deal of the controversy has to do with the isoscalar  $f_0$(500),
which in the large $N$ limit is degenerate with the $a_0$. Hence,
studying the leading $1/N^2$ corrections is presumably crucial.
Furthermore there are certain predictions of quite different
nature~\cite{Nieves:2009ez, Nieves:2011gb} that suggest that the large $N$ $a_0$ mass might coincide 
with that of the rho meson, a possibility which is not inconsistent with our results.

Altogether, it is fair to say that our continuum results are consistent with the ones by Castagnini and, 
as emphasized earlier, this is more remarkable  given the differences
in methodology of our approaches.  Even the scale-setting is different, 
since we use our own measurements of the string tension and two other methods to fix the scale.

Our next goal is to comment on other works and  results concerning the
meson spectrum at large number of colours obtained  with other
methods. A new paradigm has arisen from the AdS/CFT
correspondence~\cite{Maldacena:1997re, Gubser:1998bc, Witten:1998qj},
mapping field theory problems into others involving string theory,
supergravity or just gravity. This is, no doubt, an interesting
breakthrough introducing a new perspective in describing  some field theoretical phenomena 
and allowing the computation  of some observables. The large $N$ limit seems to be crucial
in making these new methods feasible. The original ideas concern  theories which are very
different from  QCD, being conformal, supersymmetric, and having fermions and scalars in 
the adjoint representation. As we move away from this situation the level of rigour in the 
connection decreases. Nonetheless, theories with quarks in the
fundamental~\cite{Karch:2002sh}, with reduced or
no-supersymmetry and with running couplings, have been studied. Interesting calculations 
including  meson masses have been  performed~\cite{Kruczenski:2003be,
Babington:2003vm, Kruczenski:2003uq, Sakai:2004cn}. 
We address  the reader to the review  in Ref.~\cite{Erdmenger:2007cm} for a very nice  account and a more
complete list of  references.   It is  worth mentioning that even in theories which are 
not exactly QCD some observables  become rather close to our results.
For example, the slope $Y_\rho$ obtained in 
Ref.~\cite{Babington:2003vm}  is indeed consistent with our result within errors. We should
also emphasize that more phenomenological methods inspired by
holography have been proposed~\cite{Erlich:2005qh, DaRold:2005vr,
Hirn:2005nr}. As a general rule, some of these papers can use our
results to fix some of the parameters of their models, but they have
the potential of predicting higher excited states which are more
difficult to obtain by lattice methods. Along these lines it also
worth mentioning the modified string proposal of
Ref.~\cite{Bochicchio:2016tel}  which
gives meson spectrum results which in some cases are quite compatible
with our results.

A final comment concerns possible future improvement of our work. Given the effort involved, 
the large uncertainties in taking the continuum limit are somewhat discouraging. To obtain a 
better control it is sometimes good to go to coarser lattices where the effect is more pronounced.
However, for coarser lattices using the simplest parameterization becomes more doubtful and adding
more parameters spoils the advantage. From our point of view it is better to go to finer lattices
and to reduce the errors. For that the most important limitation is the value of $N$, whose square
root translates into an effective lattice size. The present limit is only computational and  enters
in the cpu resources needed to compute the propagator. Reaching larger values allows a longer time
extent of the correlators and hence longer plateaus. Furthermore, one could reach larger values 
of $b$ and, hence, smaller values of the lattice spacing without running into  finite effective 
size problems. In relation to this, it should be mentioned that there is no need to take volume 
reduction to the extreme and simulate the 1 point lattice TEK model. The results could be achieved
by running in a small lattice of size $L^3\times L_0$. However, we advise researchers trying to follow
this road to use appropriately chosen twisted boundary conditions. In particular, using  symmetric twist
the effective lattice size would become $L\sqrt{N}$ and good results could be obtained with much smaller
values of $N$ at very small volumes. 

%% file: tables.tex
\subsection{Wilson fermions}
\label{tables_wf}

\setlength{\tabcolsep}{10pt}
\addtolength{\tabcolsep}{4pt}
\begin{table}[hp]
\small
\begin{center}
\begin{tabular}{|c|c|c|c|c|c|c|}
\hline
$N$ & $b$     &$\kappa$& $a m_{\rm pcac}$ & $\bchi$ \\
\hline
169 &0.355& 0.1592& 0.05672 (84)&  1.33 \\
169 &0.355& 0.1600& 0.04260 (99)&  1.25 \\
169 &0.355& 0.1607& 0.0306  (11)&  1.14 \\
\hline
169 &0.360& 0.1570& 0.05151 (76)&  0.68 \\
169 &0.360& 0.1577& 0.03899 (70)&  0.76 \\
169 &0.360& 0.1585& 0.02385 (82)&  0.95 \\
\hline
289 &0.355& 0.1565& 0.10781 (45)&  1.64 \\
289 &0.355& 0.1575& 0.08986 (43)&  1.50 \\
289 &0.355& 0.1585& 0.07238 (43)&  1.33 \\
289 &0.355& 0.1592& 0.06041 (43)&  1.21 \\
289 &0.355& 0.1600& 0.04695 (42)&  1.12 \\
289 &0.355& 0.1607& 0.03525 (42)&  1.12 \\
\hline
289 &0.360& 0.1550& 0.09070 (95)&  0.59 \\
289 &0.360& 0.1560& 0.07251 (82)&  0.43 \\
289 &0.360& 0.1570& 0.05467 (72)&  0.25 \\
289 &0.360& 0.1577& 0.04231 (67)&  0.17 \\
289 &0.360& 0.1585& 0.02808 (75)&  0.24 \\
\hline
289 &0.365& 0.1535& 0.0814  (11)&  0.67 \\
289 &0.365& 0.1540& 0.0720  (10)&  0.68 \\
289 &0.365& 0.1545& 0.06269 (99)&  0.70 \\
289 &0.365& 0.1550& 0.05350 (95)&  0.74 \\
289 &0.365& 0.1555& 0.04440 (90)&  0.80 \\
289 &0.365& 0.1562& 0.03187 (83)&  0.96 \\
289 &0.365& 0.1570& 0.01799 (66)&  1.19 \\
\hline
289 &0.370& 0.1520& 0.0821  (13)&  1.29 \\
289 &0.370& 0.1525& 0.0722  (12)&  1.37 \\
289 &0.370& 0.1530& 0.0624  (11)&  1.44 \\
289 &0.370& 0.1535& 0.0526  (10)&  1.50 \\
289 &0.370& 0.1540& 0.04292 (94)&  1.53 \\
289 &0.370& 0.1547& 0.02958 (84)&  1.49 \\
289 &0.370& 0.1555& 0.01481 (73)&  1.30 \\
\hline
361 &0.355& 0.1592& 0.06118 (50)&  0.37 \\
361 &0.355& 0.1600& 0.04794 (46)&  0.47 \\
361 &0.355& 0.1607& 0.03655 (43)&  0.54 \\
\hline
361 &0.360& 0.1570& 0.05631 (39)&  0.87 \\
361 &0.360& 0.1577& 0.04396 (36)&  0.88 \\
361 &0.360& 0.1585& 0.02998 (33)&  0.88 \\
\hline
361 &0.365& 0.1555& 0.04669 (77)&  0.30 \\
361 &0.365& 0.1562& 0.03374 (76)&  0.26 \\
361 &0.365& 0.1570& 0.0185  (11)&  0.23 \\
\hline
361 &0.370& 0.1540& 0.04372 (69)&  0.27 \\
361 &0.370& 0.1547& 0.03063 (60)&  0.27 \\
361 &0.370& 0.1555& 0.01601 (49)&  0.27 \\
\hline
\end{tabular}
\caption{PCAC mass for Wilson fermions.}
\label{t:mpcac_w}
\end{center}
\end{table}
\addtolength{\tabcolsep}{-4pt}

\begin{table}[t]
\small
\begin{center} 
\begin{tabular}{|c|c|c|c|c|c|c|}
\hline
$N$ & $b$     &$\kappa$&  $m_\pi/\sqrt{\sigma}$  & $\bchi$ &   $m_{\pi^*}/\sqrt{\sigma}$   &$\bchi$    \\
\hline
169&0.355& 0.1592& 1.658 (38)&  0.59 &  3.40 (08)& 0.59          \\
169&0.355& 0.1600& 1.428 (44)&  0.46 &  3.21 (07)& 0.46          \\
169&0.355& 0.1607& 1.217 (53)&  0.46 &  3.02 (11)& 0.45          \\
\hline
169&0.360& 0.1570& 1.745 (80)&  0.39 &  3.85 (16)& 0.44          \\
169&0.360& 0.1577& 1.536 (80)&  0.43 &  3.65 (15)& 0.43          \\
169&0.360& 0.1585& 1.188 (99)&  0.30 &  3.28 (12)& 0.27          \\
\hline
289&0.355& 0.1565& 2.406 (26)&  0.64 & 4.63 (30)& 0.54          \\
289&0.355& 0.1575& 2.189 (24)&  0.54 & 4.47 (23)& 0.47          \\
289&0.355& 0.1585& 1.957 (23)&  0.47 & 4.31 (20)& 0.42          \\
289&0.355& 0.1592& 1.782 (22)&  0.42 & 4.17 (20)& 0.38          \\
289&0.355& 0.1600& 1.563 (21)&  0.38 & 3.99 (20)& 0.36          \\
289&0.355& 0.1607& 1.345 (21)&  0.36 & 3.79 (19)& 0.33          \\
\hline
289&0.360& 0.1550& 2.445 (48)&  0.39 & 5.07 (31)& 0.47          \\
289&0.360& 0.1560& 2.181 (43)&  0.31 & 4.88 (28)& 0.43          \\
289&0.360& 0.1570& 1.893 (40)&  0.22 & 4.69 (25)& 0.36          \\
289&0.360& 0.1577& 1.670 (41)&  0.16 & 4.54 (26)& 0.29          \\
289&0.360& 0.1585& 1.370 (65)&  0.10 & 4.24 (43)& 0.18          \\
\hline
289&0.365& 0.1535& 2.540 (39)&  0.21 & 5.38 (14)& 0.19          \\
289&0.365& 0.1540& 2.392 (36)&  0.21 & 5.29 (14)& 0.19          \\
289&0.365& 0.1545& 2.239 (34)&  0.21 & 5.20 (14)& 0.18          \\
289&0.365& 0.1550& 2.081 (33)&  0.20 & 5.11 (14)& 0.18          \\
289&0.365& 0.1555& 1.917 (32)&  0.20 & 5.03 (14)& 0.17          \\
289&0.365& 0.1562& 1.680 (31)&  0.21 & 4.93 (15)& 0.17          \\
289&0.365& 0.1570& 1.400 (34)&  0.26 & 4.97 (25)& 0.18          \\
\hline
289&0.370& 0.1520& 2.725 (56)&  0.41 & 5.50 (16)& 0.29          \\
289&0.370& 0.1525& 2.544 (55)&  0.38 & 5.36 (15)& 0.28          \\
289&0.370& 0.1530& 2.359 (55)&  0.34 & 5.22 (16)& 0.28          \\
289&0.370& 0.1535& 2.169 (57)&  0.30 & 5.09 (17)& 0.27          \\
289&0.370& 0.1540& 1.975 (59)&  0.25 & 4.96 (18)& 0.25          \\
289&0.370& 0.1547& 1.700 (62)&  0.18 & 4.81 (20)& 0.21         \\
289&0.370& 0.1555& 1.382 (73)&  0.13 & 4.79 (26)& 0.16          \\
\hline
361&0.355& 0.1592& 1.808 (19)&  0.22 & 4.28 (17)& 0.45          \\
361&0.355& 0.1600& 1.593 (18)&  0.19 & 4.11 (16)& 0.43          \\
361&0.355& 0.1607& 1.382 (17)&  0.17 & 3.92 (16)& 0.40          \\
\hline
361&0.360& 0.1570& 1.860 (22)&  0.27 & 4.41 (14)& 0.30          \\
361&0.360& 0.1577& 1.638 (21)&  0.21 & 4.25 (14)& 0.26          \\
361&0.360& 0.1585& 1.352 (22)&  0.18 & 4.03 (15)& 0.21          \\
\hline
361&0.365& 0.1555& 1.901 (29)&  0.15 & 5.03 (11)& 0.32          \\
361&0.365& 0.1562& 1.634 (34)&  0.12 & 4.88 (13)& 0.23          \\
361&0.365& 0.1570& 1.242 (70)&  0.10 & 4.53 (29)& 0.09          \\
\hline
361&0.370& 0.1540& 2.029 (33)&  0.16 & 5.29 (12)& 0.23          \\
361&0.370& 0.1547& 1.750 (35)&  0.08 & 5.20 (14)& 0.16          \\
361&0.370& 0.1555& 1.381 (47)&  0.03 & 5.13 (21)& 0.13          \\
\hline
\end{tabular}
\caption{Wilson fermion masses in the pseudoscalar channel, extracted from a fit of the time dependence of the optimal correlator. 
The excited mass is extracted from a double exponential fit with the ground-state mass fixed, as indicated in the text.}
\label{t:mpi_w}
\end{center}
\end{table}

\begin{table}[t]
\small
\begin{center}
\begin{tabular}{|c|c|c|c|c|c|c|}
\hline
$N$ & $b$     &$\kappa$&  $m_\rho/\sqrt{\sigma}$  & $\bchi$  & $m_{\rho^*}/\sqrt{\sigma}$   &$\bchi$     \\
\hline
169 &0.355& 0.1592& 2.08 (17)&  0.98  &4.13 (0.21)& 0.92          \\ 
169 &0.355& 0.1600& 1.945 (64)&  0.98  &4.12 (0.13)& 0.89         \\ 
169 &0.355& 0.1607& 1.922 (91)&  0.76  &4.30 (0.22)& 0.65         \\ 
\hline
169 &0.360& 0.1570& 2.08 (40)&  0.52  &4.67 (0.28)& 0.36         \\ 
169 &0.360& 0.1577& 1.910 (74)&  0.49  &4.59 (0.12)& 0.35         \\ 
169 &0.360& 0.1585& 1.668 (94)&  0.43  &4.46 (0.11)& 0.32        \\ 
\hline
289 &0.355& 0.1565& 2.727 (44)&  1.04  &4.93 (0.22)& 0.88         \\ 
289 &0.355& 0.1575& 2.561 (45)&  0.98  &4.85 (0.22)& 0.84         \\ 
289 &0.355& 0.1585& 2.395 (46)&  0.91  &4.78 (0.23)& 0.78         \\ 
289 &0.355& 0.1592& 2.280 (48)&  0.85  &4.74 (0.23)& 0.73         \\ 
289 &0.355& 0.1600& 2.152 (53)&  0.76  &4.71 (0.25)& 0.65         \\ 
289 &0.355& 0.1607& 2.048 (59)&  0.68  &4.71 (0.29)& 0.60         \\ 
\hline
289 &0.360& 0.1550& 2.742 (61)&  0.50  &5.40 (0.26)& 0.48         \\ 
289 &0.360& 0.1560& 2.538 (58)&  0.51  &5.31 (0.24)& 0.51         \\ 
289 &0.360& 0.1570& 2.328 (58)&  0.51  &5.21 (0.22)& 0.52         \\ 
289 &0.360& 0.1577& 2.176 (60)&  0.50  &5.14 (0.22)& 0.50         \\ 
289 &0.360& 0.1585& 1.998 (73)&  0.45  &5.07 (0.21)& 0.41        \\ 
289 &0.365& 0.1535& 2.814 (68)&  0.45  &5.76 (0.16)& 0.35         \\ 
\hline
289 &0.365& 0.1540& 2.691 (66)&  0.47  &5.69 (0.16)& 0.36         \\ 
289 &0.365& 0.1545& 2.569 (65)&  0.47  &5.62 (0.16)& 0.36         \\ 
289 &0.365& 0.1550& 2.449 (65)&  0.47  &5.56 (0.16)& 0.37         \\ 
289 &0.365& 0.1555& 2.333 (68)&  0.46  &5.50 (0.17)& 0.36         \\ 
289 &0.365& 0.1562& 2.191 (81)&  0.40  &5.47 (0.20)& 0.32         \\ 
289 &0.365& 0.1570& 2.04 (11)&  0.31  &5.45 (0.25)& 0.25         \\ 
\hline
289 &0.370& 0.1520& 3.013 (80)&  0.34  &5.99 (0.18)& 0.26         \\ 
289 &0.370& 0.1525& 2.846 (78)&  0.35  &5.87 (0.17)& 0.28         \\ 
289 &0.370& 0.1530& 2.678 (77)&  0.37  &5.75 (0.16)& 0.31         \\ 
289 &0.370& 0.1535& 2.506 (77)&  0.38  &5.64 (0.15)& 0.32         \\ 
289 &0.370& 0.1540& 2.332 (77)&  0.38  &5.55 (0.14)& 0.34        \\ 
289 &0.370& 0.1547& 2.086 (84)&  0.36  &5.43 (0.14)& 0.34        \\ 
289 &0.370& 0.1555& 1.80 (11)&  0.31  &5.32 (0.15)& 0.30         \\ 
\hline
361 &0.355& 0.1592& 2.300 (40)&  0.42  &4.68 (0.14)& 0.53         \\ 
361 &0.355& 0.1600& 2.166 (42)&  0.47  &4.63 (0.15)& 0.56         \\ 
361 &0.355& 0.1607& 2.051 (46)&  0.48  &4.59 (0.16)& 0.58         \\ 
\hline
361 &0.360& 0.1570& 2.322 (37)&  0.55  &4.86 (0.12)& 0.58         \\ 
361 &0.360& 0.1577& 2.164 (38)&  0.62  &4.76 (0.11)& 0.63         \\ 
361 &0.360& 0.1585& 1.981 (41)&  0.72  &4.64 (0.17)& 0.70         \\ 
\hline
361 &0.365& 0.1555& 2.346 (52)&  0.25  &5.64 (0.20)& 0.29         \\ 
361 &0.365& 0.1562& 2.154 (56)&  0.26  &5.54 (0.18)& 0.24         \\ 
361 &0.365& 0.1570& 1.893 (96)&  0.36  &5.36 (0.23)& 0.26        \\ 
\hline
361 &0.370& 0.1540& 2.486 (78)&  0.16  &5.75 (0.25)& 0.37         \\ 
361 &0.370& 0.1547& 2.268 (79)&  0.12  &5.62 (0.23)& 0.35         \\ 
361 &0.370& 0.1555& 1.979 (87)&  0.15  &5.38 (0.18)& 0.43        \\ 
\hline
\end{tabular}
\caption{Masses in the vector channel for Wilson fermions. }
\label{t:mrho}
\end{center}
\end{table}

\begin{table}[t]
\small
\begin{center}
\begin{tabular}{|c|c|c|c|c|c|c|c|c|}
\hline
$N$ & $b$     &$\kappa$&  $m_{b_1}/\sqrt{\sigma}$  & $\bchi$ &  $m_{a_1}/\sqrt{\sigma}$   &$\bchi$ &  $m_{a_0}/\sqrt{\sigma}$   &$\bchi$  \\
\hline
289 &0.355& 0.1565& 4.06 (12)&  0.94 & 3.990 (88)& 0.75 & 3.807 (77)& 0.16  \\
289 &0.355& 0.1575& 3.92 (11)&  0.87 & 3.842 (87)& 0.65 & 3.622 (77)& 0.14  \\
289 &0.355& 0.1585& 3.78 (11)&  0.79 & 3.693 (87)& 0.55 & 3.425 (69)& 0.14  \\
289 &0.355& 0.1592& 3.69 (11)&  0.72 & 3.587 (88)& 0.48 & 3.277 (72)& 0.15  \\
289 &0.355& 0.1600& 3.59 (11)&  0.60 & 3.461 (90)& 0.41 & 3.093 (80)& 0.18  \\
289 &0.355& 0.1607& 3.52 (13)&  0.46 & 3.343 (99)& 0.34 & 2.863 (92)& 0.17  \\
\hline
289 &0.360& 0.1550& 4.04 (23)&  0.28 & 3.988 (87)& 0.33 & 3.728 (55)& 0.31  \\
289 &0.360& 0.1560& 3.83 (25)&  0.37 & 3.782 (80)& 0.41 & 3.479 (50)& 0.31  \\
289 &0.360& 0.1570& 3.65 (23)&  0.48 & 3.578 (75)& 0.47 & 3.206 (47)& 0.29  \\
289 &0.360& 0.1577& 3.52 (22)&  0.57 & 3.432 (72)& 0.53 & 2.984 (50)& 0.25  \\
289 &0.360& 0.1585& 3.32 (21)&  0.69 & 3.246 (71)& 0.66 & 2.656 (60)& 0.24  \\
\hline
289 &0.365& 0.1535& 4.03 (21)&  0.41 & 4.152 (90)& 0.42 & 3.841 (78)& 0.30  \\
289 &0.365& 0.1540& 3.90 (20)&  0.39 & 4.019 (86)& 0.41 & 3.688 (75)& 0.27  \\
289 &0.365& 0.1545& 3.76 (20)&  0.36 & 3.885 (84)& 0.40 & 3.528 (72)& 0.26  \\
289 &0.365& 0.1550& 3.63 (19)&  0.34 & 3.753 (82)& 0.38 & 3.360 (69)& 0.24  \\
289 &0.365& 0.1555& 3.51 (19)&  0.32 & 3.622 (81)& 0.37 & 3.178 (67)& 0.23  \\
289 &0.365& 0.1562& 3.36 (19)&  0.30 & 3.447 (82)& 0.34 & 2.889 (69)& 0.20  \\
289 &0.365& 0.1570& 3.27 (22)&  0.35 & 3.259 (86)& 0.33 & 2.553 (70)& 0.13  \\
\hline
289 &0.370& 0.1520& 4.25 (23)&  0.45 & 4.17 (13)& 0.37 & 3.835 (92)& 0.27  \\
289 &0.370& 0.1525& 4.12 (22)&  0.45 & 4.02 (137)& 0.38 & 3.656 (87)& 0.26  \\
289 &0.370& 0.1530& 3.99 (21)&  0.45 & 3.86 (11)& 0.38 & 3.471 (81)& 0.25  \\
289 &0.370& 0.1535& 3.86 (20)&  0.46 & 3.71 (11)& 0.39 & 3.278 (76)& 0.24  \\
289 &0.370& 0.1540& 3.74 (19)&  0.46 & 3.56 (10)& 0.39 & 3.078 (71)& 0.24  \\
289 &0.370& 0.1547& 3.59 (19)&  0.46 & 3.35 (10)& 0.41 & 2.788 (62)& 0.22  \\
289 &0.370& 0.1555& 3.48 (21)&  0.39 & 3.17 (11)& 0.41 & 2.438 (55)& 0.23  \\
\hline
\end{tabular}
\caption{Masses in the $a_0$, $a_1$ and $b_1$ channels for Wilson fermions. }
\label{t:mheavy}
\end{center}
\end{table}

\addtolength{\tabcolsep}{4pt}
\begin{table}[t]
\small
\begin{center}
\begin{tabular}{|c|c|c|c|c|}
\hline
$N$ & $b$     & $\kappa$ &   $Z_A^{-1} F_\pi/\sqrt{\sigma}$  &    $\bchi$  \\
\hline
169 &0.355& 0.1592& 0.322  (11 )&  0.72          \\
169 &0.355& 0.1600& 0.285  (13 )&  0.82          \\
169 &0.355& 0.1607& 0.248  (13 )&  0.76          \\
\hline
169 &0.360& 0.1570& 0.301  (19 )&  0.51          \\
169 &0.360& 0.1577& 0.261  (17 )&  0.63          \\ 
169 &0.360& 0.1585& 0.198  (19 )&  0.80          \\ 
\hline
289 &0.355& 0.1565& 0.4389 (69)&  0.95          \\ 
289 &0.355& 0.1575& 0.4194 (59)&  0.82          \\ 
289 &0.355& 0.1585& 0.3972 (50)&  0.69          \\ 
289 &0.355& 0.1592& 0.3796 (44)&  0.63          \\ 
289 &0.355& 0.1600& 0.3567 (39)&  0.60          \\ 
289 &0.355& 0.1607& 0.3325 (39)&  0.66          \\ 
\hline
289 &0.360& 0.1550& 0.422  (10 )&  0.46         \\ 
289 &0.360& 0.1560& 0.3982 (82)&  0.39          \\ 
289 &0.360& 0.1570& 0.3688 (67)&  0.30          \\ 
289 &0.360& 0.1577& 0.3431 (65)&  0.24          \\ 
289 &0.360& 0.1585& 0.302 (10)&  0.23    \\ 
\hline
289 &0.365& 0.1535& 0.4184 (89)&  0.36           \\ 
289 &0.365& 0.1540& 0.4041 (87)&  0.29           \\ 
289 &0.365& 0.1545& 0.3878 (87)&  0.24           \\ 
289 &0.365& 0.1550& 0.3691 (89)&  0.19           \\ 
289 &0.365& 0.1555& 0.3470 (93)&  0.17           \\ 
289 &0.365& 0.1562& 0.3081 (98)&  0.19           \\ 
289 &0.365& 0.1570& 0.2423 (97)&  0.27           \\ 
\hline
289 &0.370& 0.1520& 0.4228 (120)&  0.59          \\ 
289 &0.370& 0.1525& 0.4039 (111)&  0.58          \\ 
289 &0.370& 0.1530& 0.3823 (104)&  0.56          \\ 
289 &0.370& 0.1535& 0.3568 (100)&  0.54          \\ 
289 &0.370& 0.1540& 0.3259 (95)&  0.53           \\ 
289 &0.370& 0.1547& 0.2697 (86)&  0.51           \\ 
289 &0.370& 0.1555& 0.1780 (91)&  0.44           \\ 
\hline
361 &0.355& 0.1592& 0.3830 (46)&  0.39          \\ 
361 &0.355& 0.1600& 0.3612 (40)&  0.45          \\ 
361 &0.355& 0.1607& 0.3396 (36)&  0.53          \\ 
\hline
361 &0.360& 0.1570& 0.3793 (59)&  0.56          \\ 
361 &0.360& 0.1577& 0.3566 (53)&  0.52          \\ 
361 &0.360& 0.1585& 0.3242 (52)&  0.48          \\ 
\hline
361 &0.365& 0.1555& 0.3701 (66)&  0.23          \\ 
361 &0.365& 0.1562& 0.3345 (73)&  0.17          \\ 
361 &0.365& 0.1570& 0.2678 (135)&  0.31          \\ 
\hline
361 &0.370& 0.1540& 0.3702 (61)&  0.40          \\ 
361 &0.370& 0.1547& 0.3268 (63)&  0.38          \\ 
361 &0.370& 0.1555& 0.2483 (77)&  0.28          \\
\hline
\end{tabular}
\caption{Pion decay constant for Wilson fermions.}
\label{t:fpi_wilson}
\end{center}
\end{table}
\addtolength{\tabcolsep}{-4pt}
\FloatBarrier

\subsection{Twisted mass fermions}
\label{tables_tm}

\begin{table}[H]
\small
\begin{center}
\begin{tabular}{|c|c|c|c|c|c|c|}
\hline
$N$ & $b$     &$ 2 \kappa_c \mu $ & $ m_\pi/\sqrt{\sigma}$  &  $\bchi$ & $m_{\pi^*}/\sqrt{\sigma}$   &$\bchi$   \\
\hline
169 &0.355& 0.02477& 2.500 (32)&  0.21 & 4.62 (24)& 0.74        \\
169 &0.355& 0.01736& 2.067 (25)&  0.29 & 4.36 (21)& 0.76        \\
169 &0.355& 0.01344& 1.818 (23)&  0.47 & 4.20 (18)& 0.80        \\
169 &0.355& 0.00852& 1.477 (21)&  0.69 & 3.94 (15)& 0.86        \\
\hline
169 &0.360& 0.02081& 2.431 (42)&  0.56 & 4.67 (13)& 0.65        \\
169 &0.360& 0.01570& 2.092 (40)&  0.51 & 4.45 (13)& 0.63        \\
169 &0.360& 0.01217& 1.851 (39)&  0.44 & 4.29 (13)& 0.60        \\
169 &0.360& 0.00716& 1.497 (38)&  0.31 & 4.05 (14)& 0.51        \\
\hline
289 &0.355& 0.02477& 2.523 (18)&  0.73 & 5.24 (39)& 0.70        \\
289 &0.355& 0.01736& 2.068 (15)&  0.87 & 5.03 (32)& 0.83        \\
289 &0.355& 0.01344& 1.802 (14)&  0.84 & 4.92 (33)& 0.79        \\
289 &0.355& 0.00852& 1.424 (14)&  0.59 & 4.69 (41)& 0.56        \\
\hline
289 &0.360& 0.02081& 2.521 (38)&  0.68 & 5.14 (44)& 0.89        \\
289 &0.360& 0.01570& 2.159 (34)&  0.57 & 4.93 (31)& 0.72        \\
289 &0.360& 0.01217& 1.886 (32)&  0.49 & 4.78 (37)& 0.57        \\
289 &0.360& 0.00716& 1.448 (30)&  0.46 & 4.49 (31)& 0.40        \\
\hline
289 &0.365& 0.01814& 2.487 (37)&  0.50 & 5.33 (17)& 0.40        \\
289 &0.365& 0.01293& 2.082 (34)&  0.47 & 5.13 (17)& 0.39        \\
289 &0.365& 0.00933& 1.782 (32)&  0.50 & 5.00 (19)& 0.44        \\
289 &0.365& 0.00615& 1.497 (31)&  0.54 & 4.90 (22)& 0.52        \\
\hline
289 &0.370& 0.01552& 2.551 (59)&  0.53 & 5.57 (31)& 0.41        \\
289 &0.370& 0.01194& 2.221 (52)&  0.52 & 5.31 (26)& 0.42        \\
289 &0.370& 0.00827& 1.861 (50)&  0.49 & 5.05 (24)& 0.43        \\
289 &0.370& 0.00534& 1.555 (54)&  0.44 & 4.84 (26)& 0.42        \\
\hline
361 &0.355& 0.02477& 2.548 (21)&  0.54 & 3.17 (95)& 0.76        \\
361 &0.355& 0.01736& 2.082 (16)&  0.45 & 5.15 (29)& 0.57        \\
361 &0.355& 0.01344& 1.806 (14)&  0.37 & 4.96 (29)& 0.48        \\
361 &0.355& 0.00852& 1.411 (13)&  0.28 & 4.64 (33)& 0.35        \\
\hline
361 &0.360& 0.02081& 2.520 (25)&  0.37 & 5.43 (20)& 0.73        \\
361 &0.360& 0.01570& 2.150 (20)&  0.43 & 5.21 (19)& 0.72        \\
361 &0.360& 0.01217& 1.871 (18)&  0.44 & 5.07 (18)& 0.68        \\
361 &0.360& 0.00716& 1.421 (14)&  0.36 & 4.91 (20)& 0.49        \\
\hline
361 &0.365& 0.01814& 2.503 (24)&  0.83 & 5.30 (15)& 0.61        \\
361 &0.365& 0.01293& 2.076 (19)&  0.74 & 5.02 (12)& 0.55        \\
361 &0.365& 0.00933& 1.758 (17)&  0.60 & 4.83 (10)& 0.45        \\
361 &0.365& 0.00615& 1.456 (18)&  0.41 & 4.69 (10)& 0.33        \\
\hline
361 &0.370& 0.01552& 2.500 (40)&  0.41 & 5.56 (22)& 0.41        \\
361 &0.370& 0.01194& 2.183 (33)&  0.31 & 5.41 (21)& 0.34        \\
361 &0.370& 0.00827& 1.835 (29)&  0.24 & 5.31 (21)& 0.26        \\
361 &0.370& 0.00534& 1.530 (32)&  0.24 & 5.26 (25)& 0.21        \\
\hline
\end{tabular}
\caption{Masses in the pseudoscalar channel for twisted mass fermions. 
 }
\label{t:mpi_tw}
\end{center}
\end{table}

\addtolength{\tabcolsep}{4pt}
\begin{table}[t]
\small
\begin{center}
\begin{tabular}{|c|c|c|c|c|}
\hline
$N$ & $b$     & $ 2 \kappa_c \mu$ &   $F_\pi/\sqrt{\sigma}$  &    $\bchi$  \\
\hline
169 &0.355& 0.02477& 0.3644 (32)&  0.78         \\                    
169 &0.355& 0.01736& 0.3171 (19)&  0.77         \\ 
169 &0.355& 0.01344& 0.2870 (17)&  0.78         \\ 
169 &0.355& 0.00852& 0.2376 (19)&  0.81         \\ 
\hline
169 &0.360& 0.02081& 0.3425 (25)&  0.99         \\  
169 &0.360& 0.01570& 0.3058 (19)&  0.85         \\
169 &0.360& 0.01217& 0.2744 (23)&  0.72         \\  
169 &0.360& 0.00716& 0.2116 (33)&  0.52         \\
\hline
289 &0.355& 0.02477& 0.3775 (11)&  0.91         \\  
289 &0.355& 0.01736& 0.3338 (13)&  0.92         \\
289 &0.355& 0.01344& 0.3080 (14)&  0.82         \\ 
289 &0.355& 0.00852& 0.2695 (19)&  0.58         \\  
\hline
289 &0.360& 0.02081& 0.3652 (27)&  0.62         \\  
289 &0.360& 0.01570& 0.3305 (20)&  0.54         \\  
289 &0.360& 0.01217& 0.3032 (20)&  0.49         \\
289 &0.360& 0.00716& 0.2542 (32)&  0.45         \\ 
\hline
289 &0.365& 0.01814& 0.3537 (24)&  0.75         \\  
289 &0.365& 0.01293& 0.3127 (20)&  0.70         \\
289 &0.365& 0.00933& 0.2761 (24)&  0.64         \\ 
289 &0.365& 0.00615& 0.2311 (32)&  0.57         \\ 
\hline
289 &0.370& 0.01552& 0.3387 (40)&  0.57         \\  
289 &0.370& 0.01194& 0.3030 (30)&  0.54         \\
289 &0.370& 0.00827& 0.2573 (39)&  0.48         \\  
289 &0.370& 0.00534& 0.2068 (55)&  0.40         \\
\hline
361 &0.355& 0.02477& 0.3812 (14)&  0.46         \\  
361 &0.355& 0.01736& 0.3363 (12)&  0.44         \\
361 &0.355& 0.01344& 0.3106 (14)&  0.41         \\
361 &0.355& 0.00852& 0.2748 (19)&  0.38         \\ 
\hline
361 &0.360& 0.02081& 0.3708 (20)&  0.30         \\  
361 &0.360& 0.01570& 0.3372 (14)&  0.31         \\ 
361 &0.360& 0.01217& 0.3116 (15)&  0.31         \\
361 &0.360& 0.00716& 0.2681 (21)&  0.24         \\
\hline
361 &0.365& 0.01814& 0.3558 (25)&  1.03         \\  
361 &0.365& 0.01293& 0.3161 (19)&  0.91         \\
361 &0.365& 0.00933& 0.2825 (23)&  0.71         \\ 
361 &0.365& 0.00615& 0.2424 (33)&  0.45         \\
\hline
361 &0.370& 0.01552& 0.3518 (20)&  0.46         \\
361 &0.370& 0.01194& 0.3184 (16)&  0.39         \\
361 &0.370& 0.00827& 0.2752 (24)&  0.27         \\
361 &0.370& 0.00534& 0.2277 (37)&  0.20         \\
\hline
\end{tabular}
\caption{Pion decay constant for twisted mass fermions.}
\label{t:fpi_tw}
\end{center}
\end{table}
\addtolength{\tabcolsep}{-4pt}

%% file: spectrum_NF0.bbl
\providecommand{\href}[2]{#2}\begingroup\raggedright\endgroup